\documentclass[superscriptaddress,twocolumn,showpacs,preprintnumbers,amsmath,amssymb,prb]{revtex4}
\usepackage{graphicx,times,epsfig,color}
\usepackage[all]{xy}

\begin{document}

\def\salto{\vskip 1cm}
\def\lgr{\langle\langle}
\def\rgr{\rangle\rangle}

\title{A Novel Approach to Study Highly Correlated Nanostructures: \\
The Logarithmic Discretization Embedded Cluster Approximation}
%
\author{E. V. Anda}
\affiliation{Departamento de F\'{\i}sica, Pontif\'{\i}cia
Universidade Cat\'olica do Rio de Janeiro, 22453-900, Brazil}
\author{G. Chiappe}
\affiliation{Departamento de F\'{\i}sica Aplicada, Universidad de
Alicante, San Vicente del Raspeig, Alicante 03690 Spain}
\affiliation{Departamento de F\'{\i}sica J. J. Giambiagi, Facultad
de Ciencias Exactas, Universidad de Buenos Aires, Ciudad
Universitaria, 1428 Buenos Aires, Argentina}
\author{C.A. B\"usser}
\affiliation{Department of Physics and Astronomy, Ohio University,
 Athens, Ohio 45701, USA}
\author{M.A. Davidovich}
\affiliation{Departamento de F\'{\i}sica, Pontif\'{\i}cia
Universidade Cat\'olica do Rio de Janeiro, 22453-900, Brazil}
\author{G. B. Martins}
\email[Corresponding author: ]{martins@oakland.edu}
\affiliation{Department of Physics, Oakland University, Rochester,
MI 48309, USA}
\author{F. Heidrich-Meisner}
\affiliation{Institut f\"ur Theoretische Physik C, RWTH Aachen University,  52056
 Aachen, Germany}
 \affiliation{Materials  Science and Technology Division, Oak Ridge
 National Laboratory,
 Oak Ridge, Tennessee 37831, USA and\\
 Department of Physics and Astronomy, University of Tennessee,
 Knoxville,
 Tennessee 37996, USA}
\author{E. Dagotto}
\affiliation{Materials  Science and Technology Division, Oak Ridge
 National Laboratory,
 Oak Ridge, Tennessee 37831, USA and\\
 Department of Physics and Astronomy, University of Tennessee,
 Knoxville,
 Tennessee 37996, USA}


\begin{abstract}
This work proposes a new approach to study transport properties of
highly correlated local structures. The method, dubbed the
Logarithmic Discretization Embedded Cluster Approximation (LDECA),
consists of diagonalizing a finite cluster containing the many-body
terms of the Hamiltonian and embedding it into the rest of the
system, combined with Wilson's idea of a logarithmic discretization
of the representation of the Hamiltonian. The physics associated
with both one embedded dot and a double-dot side-coupled to
leads is discussed in detail. In the former case, the results
perfectly agree with Bethe ansatz data, while in the latter,
the physics obtained is framed in the conceptual background of a two-stage
Kondo problem. A many-body formalism provides a solid
theoretical foundation to the method. We argue that LDECA is well
suited to study complicated problems such as transport through molecules or
quantum dot structures with complex ground states.
\end{abstract}
\maketitle


\section{Introduction}

The study of nanostructures has been motivated, on the one hand, by the potential applications in
molecular electronics devices \cite{molecular} or in quantum computing
\cite{qcomp} and, on the other hand, by the search for a more profound understanding
of fundamental many-body physics such as the Kondo effect. Experimentally, not only the existence of the Kondo
effect in quantum dots \cite{GG} or single-molecule transistors
\cite{molecules} has been established, but it has also been
demonstrated that nanostructures can be designed to produce more
exotic phases such as multi-channel physics and thus, non-Fermi
liquid behavior.\cite{nonfermi} On the theoretical side, while the
single-impurity case is well understood by means of firmly established
analytical \cite{bethe} and numerical methods, such as
 the Numerical Renormalization Group technique (NRG),\cite{Wilson}
the search for unconventional effects, non-equilibrium behavior, and
the need to model complex real structures, such as molecules or
multi-dot geometries, has triggered the development of alternative
methods.\cite{tDMRG,fRG,ferrari99,davidovich02,anda02,chiappe03} For instance, the procedure of
exactly diagonalizing a finite cluster containing the many-body
terms and embedding it into the rest of the system, the Embedded
Cluster Approximation (ECA), has satisfactorily been used
 to study transport in nanoscopic structures in the
last few years.\cite{ferrari99,davidovich02,anda02,chiappe03}
Ideas similar to the embedded cluster method have been applied to the
metal-insulator transition of the Hubbard model.\cite{busser2, chiappe99,Tremblay}  

Incorporating ideas from the Density Matrix Renormalization Group
 method (DMRG) \cite{white}
into NRG  and vice versa has also resulted in substantial
improvements in, {\it e.g.}, the calculation of dynamical properties
\cite{hofstetter} or time-evolution schemes, \cite{anders} which now
allows one to address problems previously out of reach for either
method. In the same spirit, it is the objective of this paper to
present the Logarithmic Discretization Embedded Cluster Approximation
(LDECA) approach  to study highly correlated electrons in nano-scale systems,
combining ECA with Wilson's idea of a
logarithmic discretization of the conduction band.\cite{Wilson}
As one of our  main results, we utilize  many-body arguments to
provide a solid theoretical justification of this formalism. Although
 the ECA
method, due to the embedding process, is designed to analyze the
infinite system, it produces results that depend on the cluster's
size, which, in some cases, has led to controversial results.
\cite{busser04,hm08}
LDECA not only successfully reduces these finite-size effects, 
but, more importantly, it also optimizes the description of the system in the vicinity 
of the Fermi level, allowing for the analysis of lower energy scales than accessible to ECA. 

To demonstrate the potential of the method, we focus on the physics
of the Kondo effect in a single-dot, where we find excellent
agreement with exact Bethe ansatz (BA) results. As there is a timely interest
in more involved versions of Kondo physics, such as multi-channel
situations, \cite{nonfermi} SU(4),\cite{su4, ecasu4} as well as two-stage
Kondo (TSK) effects,\cite{wiel,two-stage,Grempel,pedro,zitko06,zitko07} we further apply LDECA
to study a double-dot structure side-connected to leads.
This system, with a subtle TSK ground state similar to the one studied
 in Ref.~\onlinecite{Grempel}, is an important
testbed for  our approach. Our results are encouraging, and we thus
 envision the successful future application
of LDECA to more involved systems such as
molecules adsorbed at metallic surfaces \cite{molecules,quique} or
dot structures with subtle ground states.\cite{nonfermi}

The plan of the paper is as follows. We first provide a discussion of the theoretical
foundation of the method in terms of diagrammatic perturbation theory in Sec.~\ref{sec:ldeca}.
For the sake of a clear presentation, we choose to provide a  pedagogical account
of the theory, therefore the details will be given in an appendix (App.~\ref{sec:app}). 
Our results for the two systems,
one embedded dot and a two dot model, are covered in Secs.~\ref{sec:1qd} and Sec.~\ref{sec:2qd}, respectively. For both models,
we discuss the local density of states and the conductance as a function of gate potential.
We close with a summary in Sec.~\ref{sec:sum}.


\section{The ``Logarithmic Discretization Embedded Cluster Approximation'' (LDECA)}
\label{sec:ldeca}

\begin{figure}
\centerline{\epsfxsize=7.cm\epsfbox{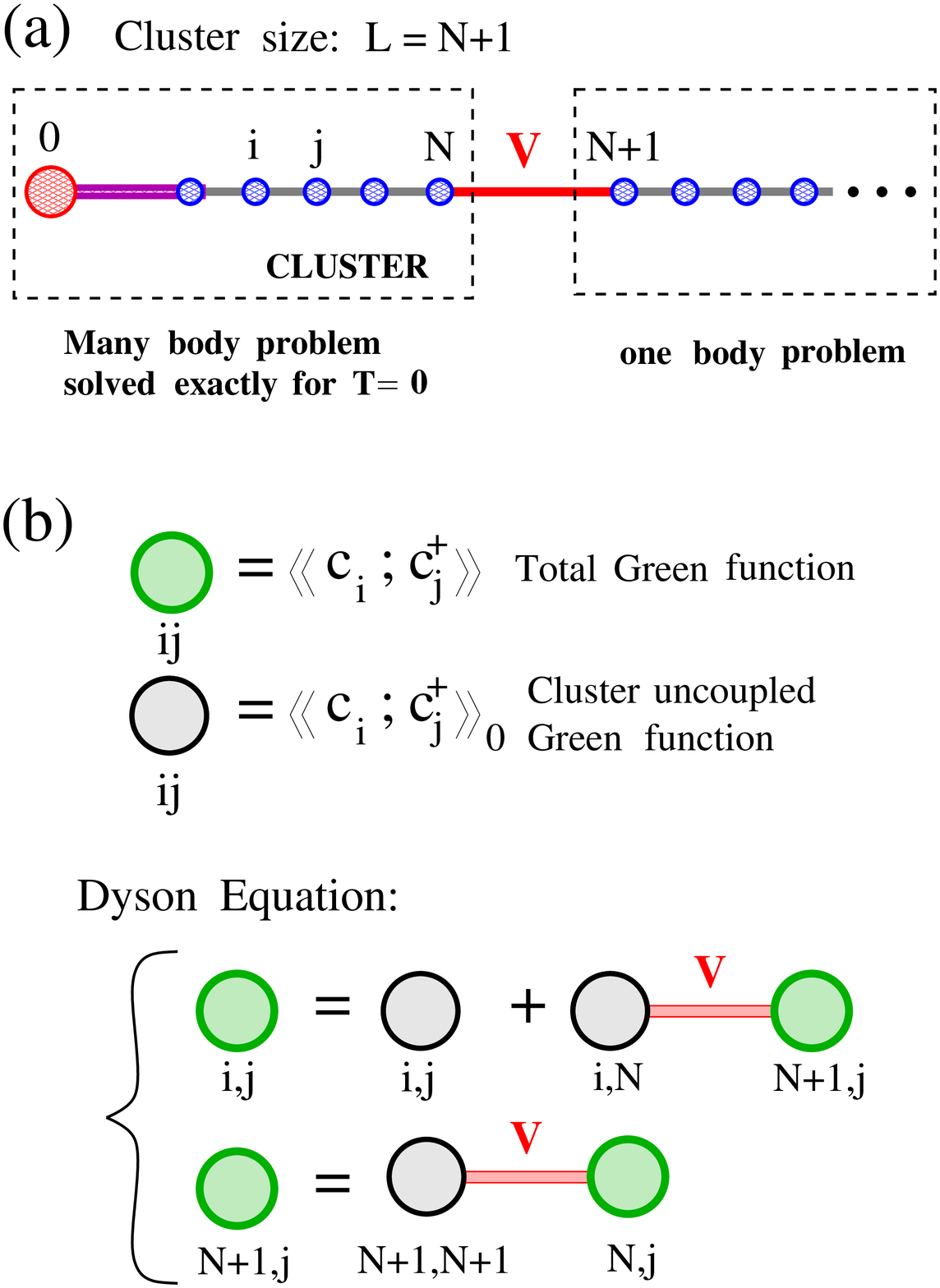}}
\caption{ (Color online) (a) Schematic description of the steps needed in the
implementation of the ECA and LDECA methods for the particular case of one
quantum dot coupled to a conduction band. The system is separated
into two parts: the first one contains the many-body terms and the first few
neighboring sites in the lead. This forms the cluster to be exactly
diagonalized, with $L=N+1$ sites. The second part is the rest of the
lead, a semi-infinite tight-binding chain. A
crucial step, which allows the simulation of Kondo physics, is the
{\it embedding} of the cluster into the remaining part
of the system. This step is performed through a Dyson
equation, which amounts to a summation of an infinite family of
Feynman diagrams arising from perturbation theory. (b) Diagrammatic
expansion associated to the Dyson equation. The dressed propagators
(in green, darkly shaded), which  reestablish the artificially broken connection
between sites N and R (through a hopping term $V$), are calculated as a function of the bare
propagators (gray, lightly shaded). Note that the equation for the dressed
propagator {\bf N+1,j} does not have an independent bare term, since the
bare propagator {\bf N+1,j} is zero.} \label{fig:Dyson}
\end{figure}

Similarly to the ECA method, LDECA is supposed to treat localized impurity systems
that consist of a region with  many-body interactions weakly coupled to
non-interacting conduction bands. The approach is based on the idea
that the many-body effects of the impurities are local in character.
With this in mind, we proceed in
three steps: first, out of the complete system, 
one isolates a cluster with $L$ sites that consists of the impurities plus their $N$
nearest neighboring sites in the tight-binding lattice (thus, $L=N+1$, in the case of a single 
impurity). In this cluster is where most of the many-body effects are expected to be confined. In a  second
step the cluster's Green function is computed with exact diagonalization, which then,
in a last step, is
{\it embedded} into the rest of the tight-binding lattice.\cite{ferrari99,davidovich02,anda02,chiappe03, ecasu4,
busser04,martins05,
interference, otras1, otras2, Armando} The
precise meaning of the embedding step is described below.

The theoretical foundation of the method is outlined using the Anderson single-impurity
Hamiltonian describing a dot connected to a semi-infinite lead. \cite{Hewsonbook}
The total Hamiltonian reads
\begin{eqnarray}
H_{T} &=& V_g \sum_{\sigma} n_{0\sigma} + H_{MB}+
\nonumber \label{Htotal} \\
&&t' \sum_\sigma (c_{0\sigma}^{\dagger} c_{1\sigma} + c_{1\sigma}^{\dagger}
c_{0\sigma}) + H_{\mbox{\small band}},
\end{eqnarray}
where
\begin{eqnarray}
H_{MB} &=& U/2 \sum_{\sigma} n_{0\sigma} n_{0\bar{\sigma}},
\end{eqnarray}
and
\begin{eqnarray}
H_{\mbox{\small band}} &=&  \sum_{i=1\sigma}^\infty t_i(c_{i\sigma}^{\dagger}
c_{i+1\sigma} + c_{i+1\sigma}^{\dagger} c_{i\sigma})\,.
\end{eqnarray}
The first two terms of $H_T$ represent the impurity, which has a
diagonal energy, the gate potential $V_g$, and a Coulomb repulsion $U$ in the Hamiltonian
$H_{MB}$. The third term is the hybridization of the impurity with
the band and finally, $H_{\mbox{\small band}}$ represents the
continuous spectrum, in this case modeled by a semi-infinite non-interacting chain.
 $c_{i\sigma}^{\dagger}$ is a fermion creation operator acting on site $i$, with a spin index $\sigma=\uparrow\downarrow$.
$n_{i\sigma}=c_{i\sigma}^{\dagger}c_{i\sigma}^{}$ is the particle density operator.  $t'$ and $t_i$ are the hopping matrix elements
between the dot and the leads and in the leads, respectively. 
 A tight-binding band with a semi-elliptical density of states is obtained with the choice of $t_i=1$.

This problem can be treated within the framework of quantum perturbation theory.
The standard many-body perturbation theory formulation generally considers the kinetic energy
as the unperturbed Hamiltonian and the many-body terms as the perturbation. This permits
the use of
Wick's theorem to formulate a diagrammatic expansion for the propagators of the system. In
our case, however, we adopt an opposite point of view. The unperturbed
Hamiltonian consists of two parts, the isolated cluster, which
includes the impurity and its neighborhood, and the rest of the
system, as represented by the two dashed boxes in Fig.~\ref{fig:Dyson}~(a).
The kinetic energy associated to the connection of
these two subsystems is now considered to be the perturbation.
This seems to be an appropriate starting point to describe a system
where the many-body interactions are local, so that the cluster may contain
most of the relevant
physics we wish to describe. However, one faces several difficulties
in a theory of this kind. The most important one is the
fact that, in this case, Wick's theorem is not valid and, as a
consequence, it cannot be used to develop the diagrammatic
expansion. However, perturbation theory provides us with a way of
proposing a locator-propagator diagrammatic expansion and
establishing a criterion to sum up the most important families of
diagrams (for details, see Appendix A).

Therefore, following the strategy outlined above,
the unperturbed Hamiltonian $H_0$ is given by:
\begin{eqnarray}
H_0 &=& H_{\mbox{\small cluster}} + H_{\mbox{\small rest}}
\label{Hzero}
\end{eqnarray}
where
\begin{eqnarray}
H_{\mbox{\small cluster}}& =& V_g \sum_{\sigma} n_{0\sigma} +
H_{MB} +  \\
&&t' \sum_\sigma c_{0\sigma}^{\dagger} c_{1\sigma}
+  \sum_{i=1\sigma}^N  t_ic_{i\sigma}^{\dagger} c_{i+1\sigma} + \mbox{h.c.} ,
\label{Hcluster}
\end{eqnarray}
and
\begin{eqnarray}
H_{\mbox{\small rest}} &=& \sum_{i=N+1\sigma}^\infty  t _i (c_{i\sigma}^{\dagger}
c_{i+1\sigma} + \mbox{h.c.}) \,.\label{Hrest}
\end{eqnarray}

Figure~\ref{fig:Dyson}(a) schematically represents the two parts of the
system. Note that one of the internal connections of the lead,
represented by a red line, labeled with a $V=t_N$ in the figure, is
artificially broken by this procedure and the two parts of the
unperturbed Hamiltonian, $H_{\mbox{\small cluster}}$ and
$H_{\mbox{\small rest}}$, can be solved exactly.
The ground state of $H_{\mbox{\small cluster}}$ with $N$
sites of the lead plus the impurity is obtained by using the Lanczos
method.\cite{Elbio} In addition, using a continued fraction scheme,
the cluster Green functions at zero
temperature are then evaluated. The Green functions for $H_{\mbox{\small rest}}$
are calculated exactly since it constitutes a one-body problem.

To restore the artificially broken connection between sites $N$
and $N+1$, the interaction between the cluster and the
rest of the lead,
\begin{eqnarray}
H_{\mbox{\small p}} &=& V \sum_\sigma c_{N\sigma}^{\dagger}
c_{N+1\sigma}+ \mbox{h.c.},
\label{eq6a}
\end{eqnarray}
is taken as the perturbation in the many-body diagrammatic expansion
for the Green functions. This step represents the embedding of the
cluster into the rest of the system.

For the sake of clarity, we restrict the discussion to the local
diagonal Green function at the impurity site, while it is straightforward to calculate a
non-diagonal Green function at two arbitrary sites $i$ and $j$  following the same prescriptions. 
To obtain
the causal Green functions we follow the standard framework of an expansion
in terms of Feynman
diagrams.\cite{Abrikosov} The
causal Green function for the impurity site can be obtained from
\begin{equation}
G_{00,\sigma}(t-t')=\frac{\left\langle
{\mathcal{T}}\{c_{0\sigma}(t)c^{\dagger}_{0\sigma}(t')S(\infty)\}\right\rangle_0}{\left\langle
S(\infty)\right\rangle_0}\,, \label{eq1T}
\end{equation}
where, as usual, $S(\infty)$ is the evolution operator and $\mathcal{T}$ is
the time order operator.  The mean values are calculated in the
ground state of the unperturbed Hamiltonian $H_0$.

The evolution operator $S(\infty)$ is expanded in increasing orders
of $H_{\mbox{\small p}}$, which, when replaced in Eq.~(\ref{eq1T}), gives rise to a
perturbation series for the Green function. 

The Green function of the system at the impurity, as discussed in the appendix,
can be written using a  general Dyson equation, as schematically shown in Fig.~\ref{fig:Dyson}(b):
\begin{equation}
G_{00,\sigma}(\omega)=G^{(0)}_{00,\sigma}(\omega)+\sum_{i}
G^{(0)}_{0i,\sigma}(\omega)\Sigma^\sigma_{i}(\omega)G_{i0,\sigma}
(\omega) \label{eq18bT}
\end{equation}
where $i$ is restricted to be either $0$ or $N$, $\omega$ denotes frequency, and the self-energy
$\Sigma^\sigma_{i}(\omega)$ is defined as
\begin{equation}
\Sigma^\sigma_{i}(\omega)=\Sigma^\sigma_{N}(\omega)\delta_{iN}+\Sigma^\sigma_{0}(\omega)\delta_{i0}.
\label{eq18cT}
\end{equation}
While $\Sigma^\sigma_{N}(\omega)$ is a simple self-energy,
$\Sigma^\sigma_{0}(\omega)$ represents an infinite expansion 
(see Eq.~(\ref{eq18a}) in the Appendix). It can only be calculated approximately, although this can be done in a
systematic way by including  terms in the expansion up to a certain
order in $U$. Diagrams with a similar topological structure appear in
the calculation of the one particle Green function for the Hubbard
or Anderson impurity Hamiltonians treated in the thermodynamic limit.\cite{qual}
In addition, in Ref.~\onlinecite{Tremblay}, a diagrammatic expansion for an interacting lattice in the strong coupling
limit was used in order to include  effects of long-range interactions
beyond the  exact diagonalization of a finite cluster.

 The key approximation of LDECA is guided by 
a comparison of the two contributions to the self-energy given in Eq.~(\ref{eq18cT}).
 While $\Sigma^\sigma_{0}$ strongly depends on the size
of the cluster through the non-diagonal Green function
$\lbrack G^{(0)}_{0N,\sigma}(\omega)\rbrack^2$, $\Sigma^\sigma_{N}$ does not. This fact can
be of great help in establishing a hierarchy between these two
contributions to the self-energy. In order to achieve this,
the applicability of this expansion is restricted to the vicinity of the Fermi
energy, where we know the physics of the Kondo regime is contained.
Following Wilson's logarithmic discretization of the lead's density
of states,\cite{Wilson} the Hamiltonian is rewritten by adopting
hopping elements that depend on the site index $i$:
\begin{equation}
t_i=\frac{(1+\lambda^{-1})}{2\lambda^{(i-1)/2}}~t \label{eq19}
\end{equation}
where $\lambda>1$ is a constant, $i\geq 1$ ($i = 0$ being
the position of the impurity), and we take $t=1$ as the unit of energy.
Note that in the limit of $\lambda \to 1$, the above expression for
$t_i$ describes a semi-elliptical band, rather than the flat band
commonly used in standard NRG calculations; however, close to the
Fermi energy the two bands have the same low-energy physics.
It is worth noting here that the above discussion applies to both
ECA and LDECA, with the exception that in ECA $\lambda$ is taken
to be 1, implying that the band is not discretized.

The implications of this logarithmic discretization with respect to
the contribution of the Hilbert space states are twofold: (i) near
the dot, states of all energies are taken into account; (ii) of the
states far from the dot, only those near the Fermi level are
considered, while high energy ones are neglected. \cite{hewson2} 
Although by this procedure high energy scales are 
not well treated, it permits to accurately describe 
much smaller energy scales than it is possible with $\lambda=1$, 
for the same cluster size. This
procedure is justified if the physics of the problem
depends only on states with energy close to the Fermi level, as it
is the case in the Kondo effect, discussed in this paper.

As discussed in the appendix, around Eq.~(\ref{eq20}), from Eqs.~(\ref{eq18a1}) and (\ref{eq18a}) one realizes
that $\Sigma^\sigma_{N}
\sim \lambda^{-(N-1)}$ and $\Sigma^\sigma_{0} \sim f(N) ~ \lambda^{-(N-1)N/2}$,
such that
\begin{equation}
\frac{\Sigma^\sigma_{0}(\omega)}{\Sigma^\sigma_{N}(\omega)}  \sim f(N) \,\lambda^{-(N-1)(N/2-1)}\,.
\label{eq20T}
\end{equation}
The function $f(N)$ is an intricate function of N which goes asymptotically
to zero as N increases above the Kondo cloud length $\xi_K$, wich has a magnitude 
inversely proportional to the Kondo temperature\cite{Affleck}. This is 
the reason why even for $\lambda =1$ (ECA), the self energy 
$\Sigma^\sigma_{0}$ will eventually become negligible in the limit of $N\gg \xi_K$ 
and can be disregarded for sufficiently large N. However, this is a 
length scale typically much larger than the value for which, according to 
Eq. (\ref{eq20T}), the self energy $\Sigma^\sigma_{0}$ can be neglected in comparison to $\Sigma^\sigma_{N}$. 
Taking $\lambda=\sqrt{2}$, for example, for a cluster with $N=9$, 
we get $\Sigma^\sigma_{0}/\Sigma^\sigma_{N} \sim 10^{-5}$, reflecting
the fact that, for a cluster size which permits diagonalization with a
relatively modest numerical effort, the contribution to the self-energy can be reduced to $\Sigma^\sigma_{N}$.
This is a very favorable situation because $\Sigma^\sigma_{0}$ is a very complex object 
(see Appendix) that can be obtained only approximately, while $\Sigma^\sigma_{N}$ is very simple 
and can be calculated exactly.
Therefore, here lies the key reason to introduce the logarithmic discretization into the procedure. 

Within  this approximation, {\it i.e.}, neglecting $\Sigma^\sigma_{0}$, the embedding is carried out using Eq.~(\ref{eq18a1}) 
and therefore, Eq.~(\ref{eq18bT}) can be simplified to 
\begin{equation}
G_{00\sigma}(\omega)=G^{(0)}_{00\sigma}(\omega)+
G^{(0)}_{0N\sigma}V^2g_{N+1,\sigma}(\omega)G_{N0,\sigma} (\omega)\,.
\label{finalform}
\end{equation}

Note that the Green function of the semi-infinite linear chain at the site
N+1, $g_{N+1,\sigma}(\omega)$, representing the leads, depends on
the value of the parameter $\lambda$. For $\lambda=1$, it is the Green
function, $g^{sc}$, of a uniform semi-linear chain, given by
$g^{sc}(\omega) = (\omega \pm \sqrt{\omega^2 - 4t^2})/(2t^2)$.

Before presenting results obtained with LDECA in the next
two sections, we want to discuss some general aspects of the
embedding procedure. If we were to study the low energy excitations
by diagonalizing an undressed cluster {\it without} performing the
embedding, the necessity of incorporating a large amount of states
lying in the Kondo peak region would require the diagonalization of
a cluster of $N$ sites such that $\lambda^{-N/2}t\lesssim T_{\rm k}$, 
{\it i.e.}, the energy scale associated with the  broken link $V$ in 
Fig.~\ref{fig:Dyson}(a) would have to be less than the 
Kondo temperature. In order to fulfill this condition, and 
at the same time choose a value of $\lambda$ that still 
adequately describes the neighborhood of the Fermi energy, 
the value of $N$ would have to be such that the Lanczos diagonalization 
would become impractical due to the size of the Hilbert space.
The embedding process solves this problem in
a simple way by rendering the numerical diagonalization of a small
cluster compatible with a correct description of the 
energy region immediately around the Fermi by allowing the contribution
$\Sigma^\sigma_{0}(\omega)$ to the self-energy to be disregarded.
This approximation, even for $\lambda=1$, has shown to be
surprisingly reliable to reproduce the Kondo regime properties of
various systems, showing sufficiently fast convergence 
with cluster size.\cite{otras2, Busser1}

\begin{figure}
\includegraphics[width=2.5in]{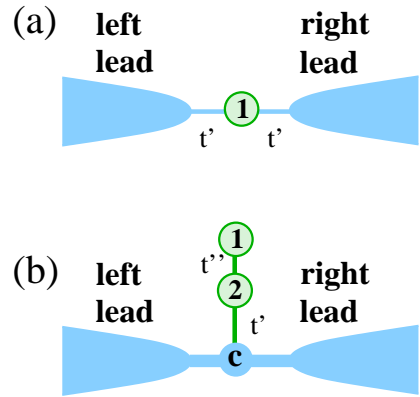}
\caption{(Color online) Sketch of the quantum dot geometries studied in this work.
(a) Single quantum dot connected to two leads. (b) Two side-connected quantum dots studied in Sec.~\ref{sec:2qd}.}
\label{fig:sketch-dots}
\end{figure}


\section{Results: LDECA and 1QD}
\label{sec:1qd}

In this section, LDECA is applied to study the conductance and the local density of states (LDOS) of a single
quantum dot connected to two leads [see Fig.\ref{fig:sketch-dots}(a)]. This case allows
for a comparison of our results with an exact solution obtained from BA.\cite{bethe}
Applying a standard basis transformation onto symmetric and
 anti-symmetric
combinations of states (at sites located symmetrically with respect
to the dot) the two-leads Hamiltonian can be effectively written
as having only one lead, rendering it identical to Eq.~(\ref{Htotal}).
This shows that this example constitutes a one-channel Kondo problem.\cite{conduc}

\begin{figure}
\includegraphics[width=3.4in]{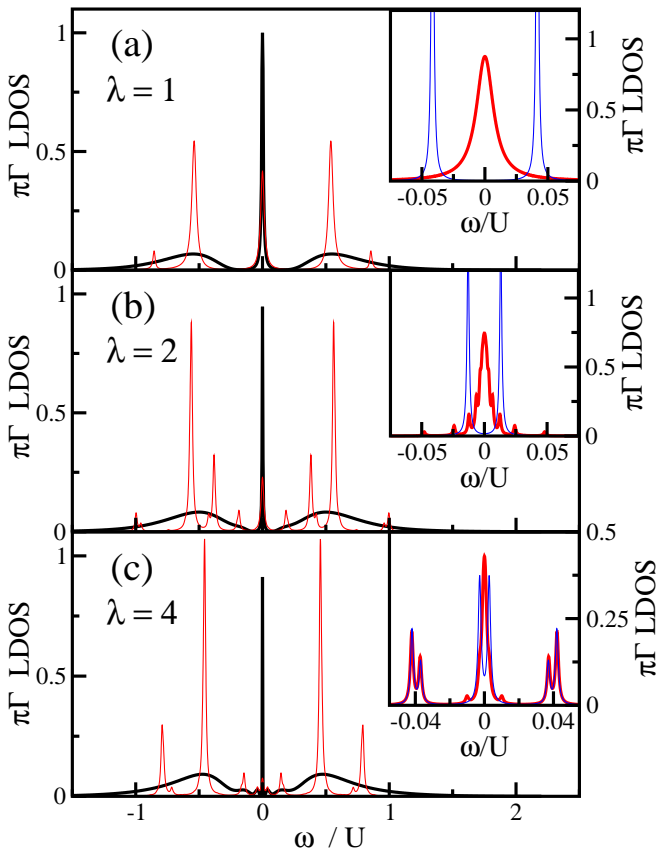}
\caption{(Color online) (a)-(c): Local density of states for  $\lambda=1,2$ and $4$, respectively,
and a cluster size $L=8$. Thin (red) lines
represent the LDOS as obtained directly  from the LDECA procedure (ECA in the case of the top panel),
while thick (black) lines are the  LDOS for the same parameters, but
after the broadening of the resonances using (Eq.~\ref{LogGauss}) has been applied, as explained in Sec.\ref{sec:ldos}.
In the insets, we show a comparison of the LDOS of the bare cluster
(thin (blue) line) and after embedding (thick (red) line) for small values of $\omega$.
All calculations done for $U=1.0$ and $U/\pi\Gamma=6.3$.}
\label{fig:DOS1qdl}
\end{figure}

\subsection{The local density of states}
\label{sec:ldos}

 We start with a discussion of the effect of the $\lambda$-discretization on the
LDOS. We expect that a larger density of poles close to the Fermi 
energy $E_{\rm F}$ is induced by the discretization, while fewer 
poles will be present away from $E_{\rm F}$. The first aspect, the accumulation of poles close to $E_{\rm F}$
is advantageous to properly describe  Kondo physics.
To still obtain a reasonable approximation to the LDOS away from the $E_{\rm F}$, 
it turns out that it is preferable to use a $\omega$-dependent broadening scheme, and we first 
detail this technical aspect.

The dressed LDOS at the impurity is a
collection of poles located at $\omega_p$, each one with its own
weight $W_p$, given by the non-linear Dyson equation used in LDECA.
As a consequence of the LDOS normalization, the weights satisfy $\sum_p W_p=2$.
In order to avoid distorting the LDOS curve through the artificially large  
separation of the poles away from the Fermi energy caused by the
logarithmic discretization, and following methods employed in NRG,\cite{bulla,Hewsonbook}
we write the LDOS as a sum of logarithmic Gaussians,\cite{bulla}
\begin{equation}
\rho(\omega) = \sum_{p}
~~\frac{{\mbox e}^{-b^2/4}}{b~W_p~\sqrt{\pi}} ~{\mbox{exp}}\left(\frac{(\ln{\omega}-\ln{W_p})^2}{b^2} \right)
\label{LogGauss}
\end{equation}
where $b$ is an arbitrary number that defines, together with $W_p$, the width of a pole
located at $\omega_p$.
We choose logarithmic Gaussians to represent the delta functions rather than the usual
Gaussians or Lorentzians because this function is asymmetric with respect to $\omega_p$.
This asymmetry, which effectively shifts the spectral weight 
of each pole to higher energies, compensates for the accumulation of poles at low energies in relation to higher
energies, caused by the logarithmic discretization.\cite{bulla}
As pointed out in Ref.~\onlinecite{Hewsonbook}, this procedure results in high-energy peaks
that are slightly broader and asymmetric than in the case of the true LDOS.

In Fig.~\ref{fig:DOS1qdl}, we show the LDOS at the impurity for
the  particle-hole symmetric situation $V_{\rm g}=-U/2$ and different
 values of $\lambda$ (thin red line) calculated with LDECA with a cluster of
$L=8$ sites. An imaginary part $\eta=0.001$, common to all poles, was
used for all curves that are obtained with a plain 
Lorentzian broadening of delta-functions (thin (red) lines in the main panels). The thick black lines show
the LDOS using the logarithmic Gaussian broadening with $b=0.5$. In this case, we obtain
 the characteristic LDOS for the Kondo problem, consisting
of a three-peak structure, with two of them located at $\omega=V_{\rm g}$ and
$\omega=V_{\rm g}+U$ and the third one, the Kondo resonance, located at
the Fermi level $E_{\rm F}=0$. The notable difference between
the LDOS for $\lambda=1$ and $\lambda>1$ is the sizeable narrowing
of the Kondo peak, in qualitative agreement with NRG \cite{Hewsonbook} (compare panels (a) with 
panels (b) and (c) in Fig.~\ref{fig:DOS1qdl}). 
A more quantitative comparison with, {\it e.g.}, NRG, will be presented elsewhere.
We wish to draw the reader's attention to the inset of each panel, showing a comparison of the undressed
LDOS [thin (blue) line] with the dressed LDOS [thick (red) line].
One can clearly see that the LDOS for the `bare' cluster (before embedding) vanishes
at the Fermi energy, while the LDOS after embedding
is finite at $\omega=0$, corroborating the notion that the embedding step is crucial
to capture Kondo physics.

\begin{figure}
\includegraphics[width=3.in]{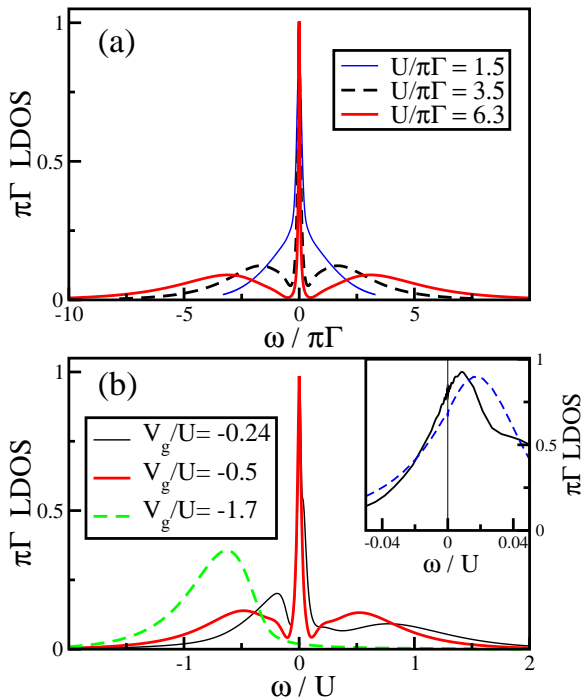}
\caption{(Color online) (a) Local density of states for the particle-hole symmetric gate
potential $V_g=-U/2$ for $U=1.0$, $\lambda=2$, and different values of $U/\pi\Gamma$, 
going from the Kondo regime ($U/\pi\Gamma=6.3$) to the intermediate valence regime ($U/\pi\Gamma=1.5$).
(b) LDOS for three different values of the gate potential for $U/\pi\Gamma=3.5$ and
$\lambda=2$. The inset shows a comparison of results for $V_g/U=-0.24$ between 
$\lambda=2$ [solid (black) curve, same as in the main panel] and $\lambda=1$ [dashed (blue) curve]. 
This illustrates the better `pinning' of the Kondo peak to the Fermi energy, 
achieved with $\lambda>1$.}
\label{fig:DOS1qd}
\end{figure}

Figure~\ref{fig:DOS1qd}(a) shows the LDOS for several values of the hybridization parameter $\Gamma$ at the
particle-hole symmetric point, and Fig.~\ref{fig:DOS1qd}(b) for a fixed ratio of 
$U/\pi\Gamma = 3.5$ and several values of the gate potential. We define the hybridization as
$\Gamma=\pi \rho_0 t'^2$,
where $\rho_0$ is the band density of states at the Fermi level. The top panel illustrates how the width of the Kondo
resonance, {\it i.e.}, $T_{\rm K}$, decreases when 
$\Gamma$ is reduced. In the bottom panel, we see how the  LDOS for a fixed
$U/\pi\Gamma = 3.5$ changes as the gate potential $V_{\rm g}$ is varied.
Note that the Kondo resonance is pinned at $E_{\rm F}$.
In contrast, for $|V_{\rm g}|>U$ [dashed (green) curve, for $V_{\rm g}/U=-1.7$] the quantum dot is
doubly occupied and therefore there is no Kondo effect. In such a case,
the LDOS has just one broad peak located at $\omega=V_{\rm g}+U$.

\subsection{The conductance}
\begin{figure}[t]
\includegraphics[width=3.in]{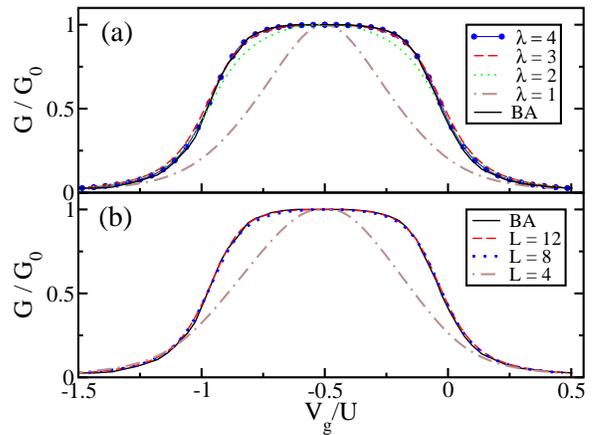}
\caption{(Color online) Comparison of LDECA and BA. Conductance vs.
$V_{\rm g}$ for $U=0.96$ and $t^{\prime}=0.2$. (a) LDECA for $L=12$ and
the values of $\lambda$ indicated. (b) LDECA for $L=4$, $8$,
and $12$, with $\lambda=4$. The BA result is the solid black line in both (a) and (b).\cite{bethe}}
\label{fig:1}
\end{figure}

Next, we demonstrate the effect of the $\lambda$-discretization on the conductance.
The conductance as a function of the gate potential $V_{\rm g}$ for a single embedded quantum dot
 is calculated by using
the Keldysh formalism.\cite{conduc} In Fig.~\ref{fig:1}~(a), we
present the conductance for a fixed cluster size ($L=12$) and different values of $\lambda$,
together with the exact value obtained by the BA. \cite{bethe} For
$\lambda=1$ (dot-dashed curve), there is a large discrepancy with the exact results:
The conductance peak is too narrow, indicating that, in this case, the
 role played by the self-energy $\Sigma^\sigma_{0}$ cannot be neglected. However, for
$\lambda=2$ (dotted curve), a value typically used in NRG calculations,\cite{Wilson} 
our LDECA results substantially improve over the ECA ones ({\it i.e.}, $\lambda=1$), 
and for  values of $\lambda$ between 3 (dashed curve) and 4 (large-dots curve),
the results accurately agree with BA. We have verified that LDECA reproduces BA results
for $U/\Gamma$ as large as $U/\Gamma=25$. 

As discussed above, the effect of neglecting the
self-energy $\Sigma^\sigma_{0}$ depends on both the value of $\lambda$ and the size of the cluster.
The dependence of the conductance on cluster size is shown in
Fig.~\ref{fig:1}~(b) for $\lambda=4$ and $L=4,8$, and $12$. Beyond a cluster size $L_c(\lambda)$, the conductance is almost
independent of $L$. For instance, for $\lambda = 4$, results for $L>8$ are
indistinguishable. 
This characteristic length $L_c(\lambda)$ decreases as $\lambda$ increases. 
As $\lambda$ controls the extension of the neighborhood of the Fermi energy that is accurately 
described, {\it i.e.}, the larger the value of $\lambda$, the smaller 
this region is, a compromise has to be found between the size of the cluster and 
the extension of the energy region around the Fermi energy that needs to be accurately 
described. Obviously, this depends on the model and the property being 
analyzed. 
The important point to be emphasized is: the results in Fig.~\ref{fig:1} show that, 
with a value of $\lambda$ similar to the one widely used by
the NRG community, it is possible to reproduce the exact results
using a cluster size accessible to the Lanczos algorithm.

It is also interesting to note that the improvement of the conductance results
for $\lambda>1$ as compared to $\lambda=1$ are associated with a better `pinning' of the Kondo
peak to the Fermi energy. This can be partially inferred from the LDOS results
shown in Fig.~\ref{fig:DOS1qd}(b), where the two solid  curves have the
Kondo peak pinned at the Fermi energy. This statement can be 
made more quantitative by considering the results in the inset of Fig.~\ref{fig:DOS1qd}(b), 
showing a comparison between $\lambda=1$ and 2. In that inset, 
the solid (black) curve is an enlarged view of the LDOS at the vicinity of 
the Fermi energy for the $V_g/U=-0.24$ curve 
presented in panel (b), which was calculated with $\lambda=2$. The dashed (blue) curve 
has been calculated with the same parameters, but for $\lambda=1$. The comparison 
clearly shows that the $\lambda=2$ result has more spectral weight at 
the Fermi energy than the $\lambda=1$ result. This increase of the spectral weight in the LDOS 
is at the heart of the improvement achieved for the conductance
by using the band discretization.


\section{Results: LDECA and a Two-Stage Kondo System}
\label{sec:2qd}

\subsection{Overview: Regimes of the model}

We next analyze a system composed of a double-dot side-connected to a
lead. The inter-dot and dot-lead connections are given by the matrix
elements $t^{\prime\prime}$ and $t^{\prime}$, respectively, as
sketched in Fig.~\ref{fig:sketch-dots}(b). The transformation to symmetric and
antisymmetric states is applied, since we again deal with a
one-channel Kondo problem.  After performing that transformation, the  Hamiltonian is given by:
\begin{eqnarray}
H_{T} &=& V_g \sum_{d,\sigma} n_{d\sigma} + U/2 \sum_{d,\sigma} n_{d\sigma} n_{d\bar{\sigma}} 
\nonumber \\
&&+t' \sqrt{2} \sum_\sigma (c_{d_{2}\sigma}^{\dagger} c_{1\sigma} + c_{1\sigma}^{\dagger}
c_{d_{2}\sigma}) + H_{\mbox{\small band}}\label{Htotal2qd}\\
H_{\mbox{\small band}} &=&  \sum_{i=1\sigma}^\infty t_i(c_{i\sigma}^{\dagger}
c_{i+1\sigma} + c_{i+1\sigma}^{\dagger} c_{i\sigma})\nonumber\,,
\end{eqnarray}
where we use the $t_i$ as given in Eq.~(\ref{eq19}), and $d=d_{1},d_{2}$, labeling dot 1 and dot 2, respectively.

The transport properties of this two-dot
system can be expected to be controlled by the interplay between the
Kondo effect and the antiferromagnetic inter-dot correlation, and by
the interference arising from the two distinct paths available to the electrons: visiting or bypassing 
the dots.\cite{Grempel} 
 
As previously discussed in the literature, systems similar to the
one depicted in Fig.~\ref{fig:sketch-dots}(b), such as, for instance, the so-called T-configuration,\cite{Grempel,zitko06}
exhibit two distinct regimes depending on the ratio $t^{\prime \prime}/t^{\prime}$:
(i) when $t^{\prime \prime} \gg t^{\prime}$, one is in the molecular
regime and (ii) for $t^{\prime \prime} \ll t^{\prime}$, the system  crosses over into the TSK regime.
It is important to realize that independently of $t^{\prime\prime}$, we expect perfect conductance
at $V_g=-U/2$. Indeed, at the particle-hole symmetric point the dots always form a singlet, which is of different nature though,
depending on the ratio $t^{\prime\prime}/t^{\prime}$, as explained below.

In the molecular regime, on the one hand, both dots act as a single entity, in a way that, as a function of the gate potential,
whenever an overall finite magnetic moment is located in the structure, 
the system exhibits a single-stage Kondo effect. In this regime of 
$t^{\prime\prime} \gg t^{\prime}$, the system essentially behaves as a single-dot 
with the two relevant levels separated by a large energy. 
On the other hand, in the limit of a small $t^{\prime\prime}/t^{\prime}$, such that the 
effective antiferromagnetic spin-spin interaction between the dots satisfies
$J$=$4(t^{\prime\prime})^2/U<T_{\rm K}$, the system is in the  two-stage Kondo regime, which is  characterized by a new 
energy scale $T_0$ related to dot~$1$, much lower than the Kondo temperature $T_{\rm K}$ 
associated to dot~$2$. For $T<T_0$, very low energy physics is involved,\cite{two-stage,Grempel} difficult to be captured by 
numerical methods such as standard ECA or DMRG. Yet, as shown below, a correct result for the conductance 
can be obtained from LDECA.

Note that TSK behavior may manifest itself both as a function of temperature
at a fixed gate potential\cite{Grempel} and as a function of gate potential
at a fixed temperature.\cite{zitko06} As our method is a zero-temperature one, we will here 
focus on the gate potential dependence of the conductance and other quantities. 
We will argue that as one starts from the empty orbital regime, first 
a single Kondo effect emerges at $V_g \gtrsim 0$,
which causes a suppression of the conductance. 
As the gate potential is further tuned towards $V_g=-U/2$, the magnetic moment of dot~1 is eventually
Kondo-screened as well through the quasi-particles of the composite system 
of dot~2 and the lead. This gives rise to the aforementioned singlet between 
dots~1 and 2, which leads to perfect conductance at $V_g=-U/2$. 

The plan of this section Sec.~\ref{sec:2qd} is thus the following. We first demonstrate 
in Sec.~\ref{sec:spin} that indeed, a singlet is formed between the two dots, independently 
of $t^{\prime\prime}/t^{\prime}$. These results are obtained with both DMRG and a diagonalization 
of the bare clusters, before the embedding process.
Second, we compute the conductance and charge as a function of gate potential 
and discuss these results both in the molecular and the TSK regimes in Secs.~\ref{sec:molecule} 
and \ref{sec:tsk}, respectively. We further aim at illustrating how the properties of 
the system change as it crosses over from the molecular regime into  the TSK regime.
Finally, at particle hole symmetry ($V_g=-U/2$), we present LDECA results for the LDOS at the dots 
and the conductance as a function of $t^{\prime\prime}/t^{\prime}$.
As a key result, we demonstrate that using the discretization of the band, LDECA 
produces perfect conductance down to very low values of $t^{\prime\prime}/t^{\prime}$, 
a result which was previously out of reach for ECA ($\lambda = 1$).


\begin{figure}[t]
\includegraphics[width=3.2in]{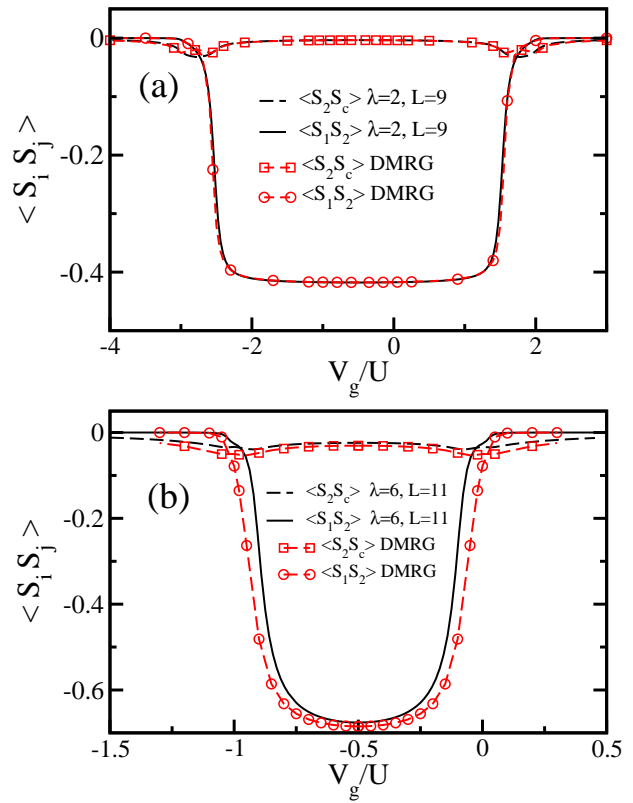}
\caption{(Color online) Spin-spin correlations calculated by
Lanczos and DMRG as a function of $V_{\rm g}$ for $U=1.0$ and $t^{\prime}=0.3$. 
(a) $t^{\prime\prime}=2.0$ (molecular regime) and (b) $t^{\prime\prime}=0.05$ (TSK regime). 
Lanczos with $L =11$ and $\lambda=6$, $\langle
{\bf S}_1\cdot {\bf S}_2\rangle$ (solid) and $\langle {\bf S}_2\cdot
{\bf S}_c\rangle$ (dashed) black lines; DMRG with $196$ sites at half filling of the full system and
$\lambda=1$, $\langle {\bf S}_1\cdot {\bf S}_2\rangle$ (circles)
and $\langle {\bf S}_2\cdot {\bf S}_c\rangle$ (squares). 
}
\label{fig:3a}
\end{figure}

\begin{figure}[t]
\includegraphics[width=3.2in]{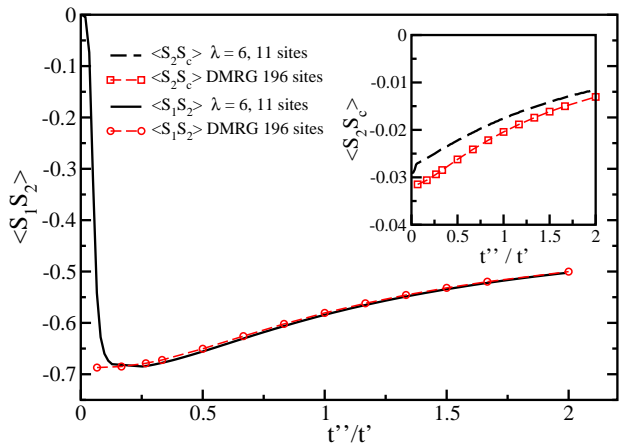}
\caption{(Color online) Spin-spin correlations
as a function of $t^{\prime\prime}/ t^{\prime}$ for $U=1.0$, $t^{\prime}=0.3$
and $V_{\rm g}=-U/2$. See legend and Fig.~\ref{fig:3a} for the description of the lines.}
\label{fig:3b}
\end{figure}

\subsection{Spin-spin correlations}
\label{sec:spin}

For the present model, we now establish the presence of a strong antiferromagnetic correlation between the dots
by analyzing the
spin-spin correlations as a function of $V_{\rm g}$, presented in Fig.~\ref{fig:3a} for both $t^{\prime\prime}=2.0$ 
(upper panel) and $t^{\prime\prime}=0.05$ (lower panel), with $t^{\prime}=0.3$ in both cases. 
Results are for large clusters with $196$ sites and $\lambda=1$,
obtained with DMRG, and also for $L=9$ and $\lambda=2$ (upper panel), and for $L=11$ and $\lambda=6$ 
(lower panel), using a Lanczos diagonalization procedure. 
Both DMRG and Lanczos calculations were done {\it without} embedding. 

In the case of the molecular regime [$t''=2$, Fig.~\ref{fig:3a}(a)], the inter-dot 
correlation  $\langle {\bf S}_1\cdot {\bf S}_2 \rangle$ 
is large for $ -2.5\lesssim V_g/U \lesssim 1.5$. Therefore, we expect a perfect conductance 
in that window, and a Kondo anti-resonance to appear at $V_g/U\gtrsim 1.5$ (see Fig.~\ref{fig:DOS-2qdA}(a), below).
For the smaller  $t^{\prime\prime}=0.05$ [Fig.~\ref{fig:3a}(b)], the antiferromagnetic correlation 
between the dots is dominant in the window $-U <V_{\rm g}<0$, indicating the formation of a singlet.
While the inter-dot spin correlation $\langle {\bf S}_1\cdot {\bf S}_2 \rangle$ is large at the
electron-hole symmetric point $V_{\rm g}=-U/2$, the antiferromagnetic
 correlation $\langle
{\bf S}_2\cdot {\bf S}_c\rangle$, although small, is not zero. Note that site
{\it c} is adjacent to dot~2, see Fig.~\ref{fig:sketch-dots}(b). 
For instance, $\langle
{\bf S}_2\cdot {\bf S}_c\rangle$ takes a maximum at $V_g\approx 0$ in the case of $t^{\prime\prime}=0.05$,
indicative of the single-stage Kondo effect that is observed in that gate potential region 
[see Fig.~\ref{fig:DOS-2qdB}(a)]. 

We now study both correlations as a function of $t^{\prime\prime}/t^{\prime}$ at $V_g=-U/2$, which is displayed in Fig.~\ref{fig:3b}.
An important observation is that $\langle {\bf S}_2\cdot {\bf S}_c\rangle$ {\it increases} in magnitude as $t^{\prime\prime}$ is {\it
 reduced}, as is shown in the inset of Fig.~\ref{fig:3b}.
This fact indicates the subtle existence of a Kondo-like ground state,
which is strengthened when $t^{\prime\prime}$ is reduced.
In addition, the antiferromagnetic inter-dot correlation, presented in the main panel of Fig.~\ref{fig:3b}, also increases
 when $t^{\prime\prime}$ is reduced,
taking values as large as $\langle {\bf S}_1\cdot {\bf S}_2\rangle
 \approx -0.7$, for $t^{\prime\prime}/t'\gtrsim 1/10$.
This is surprising, since the interplay of these two correlations
has a  behavior opposite to other known systems, such as heavy
fermions near a quantum  phase transition \cite{Coqblin} or embedded
two-dot nanostructures.\cite{Busser1} It reflects the existence of
an inter-dot singlet for all values of $t^{\prime\prime}$$>$0. However,
the nature of the singlet for small $t^{\prime\prime}/t^{\prime}$ is
different from that for the large $t^{\prime\prime}/t^{\prime}$ regime.
While in the latter case the singlet, which is caused by the direct interaction between the dots, 
destroys the Kondo regime, in the former case it is enhanced by the Kondo spin correlation with the
intervention of the conduction electrons, as shown by the fact that the inter-dot {\it and} the Kondo
 spin correlations increase when $t^{\prime\prime}$ is reduced.
These results illustrate the characteristics of a TSK state. The comparison of
data from large clusters ($196$ sites, $\lambda=1$) and short ones
($L=11$, $\lambda =6$) yields a convincing agreement for the spin
correlations, especially for $\langle{\bf S}_1\cdot {\bf S}_2\rangle$, governing the inter-dot singlet.
This agreement indicates that the spin-spin correlations are, so to speak, localized objects. 
The embedding process is not as important to calculate static properties as it is for the
conductance, which we shall see later. Still, the vanishing
of $\langle {\bf S}_1\cdot {\bf S}_2\rangle$ at very small
$t^{\prime\prime}/t^{\prime}$  on the smaller cluster reflects that,
in this particular limit, the embedding is crucial to overcome this finite-size effect.
An important point that we want to emphasize here is that the molecular and TSK results suggest
perfect conductance at $V_{\rm g}=-U/2$ for any nonzero $t^{\prime\prime}$.

\begin{figure}[t]
\vspace{3mm} 
\includegraphics[width=3.4in]{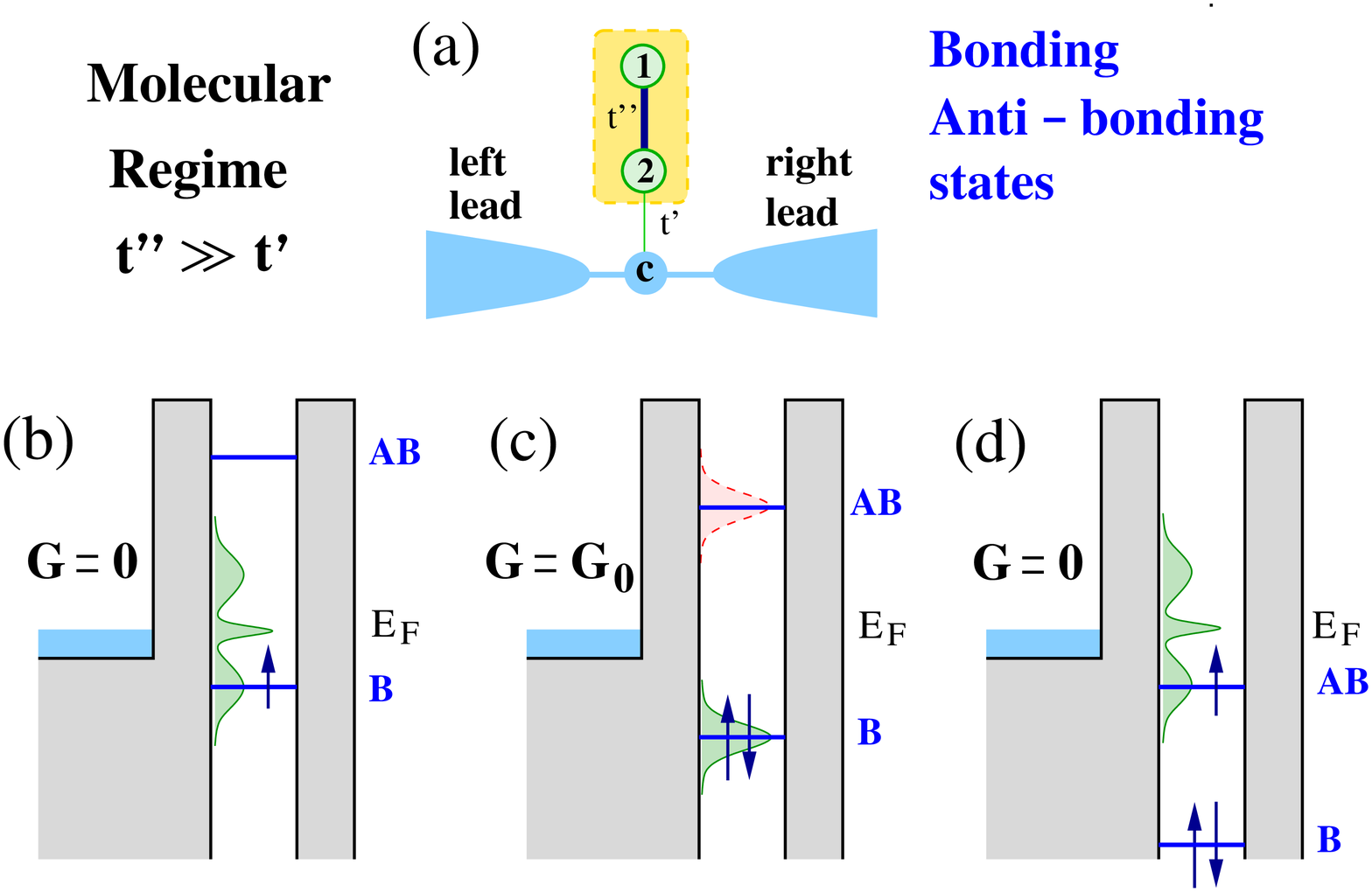}
\caption{(Color online) Schematic representation of the molecular regime. 
(a) When $t^{\prime \prime} \gg t^{\prime}$, the two dots form a single entity, 
represented by the shaded box. (b) - (d) Different regimes at different values
of gate potential: (b) Kondo effect for the bonding orbital (zero 
conductance, because of the back-scattering density of states at the Fermi energy $E_{\rm F}$), 
(c) Perfect conductance at $V_g=-U/2$ (no back-scattering density of states at 
$E_{\rm F}$), (d) Kondo effect for the anti-bonding orbital (zero conductance).}
\label{cartoon-mol}
\end{figure}


\subsection{Molecular regime}
\label{sec:molecule}

We proceed with  an analysis of  the conductance for 
the so-called `molecular' regime. 
In Fig.~\ref{cartoon-mol}(a), we schematically illustrate what happens for
$t^{\prime \prime}/t^{\prime} \gg 1.0$, {\it i.e.}, when
the independent dots are `locked' into `molecular' bonding and anti-bonding orbitals, 
separated by a large energy, proportional to $t^{\prime \prime}$. 
The two dots now behave as a single structure, represented by the dashed 
square box, side-connected to the leads. The effect of these molecular orbitals over 
the conductance through the leads, as the gate potential $V_g$ varies, is pictured in panels (b) to (d), 
where now the bonding and anti-bonding orbitals are depicted inside a quantum well. 
In panel (b), the bonding orbital is in the Kondo regime. 
As the double-dot structure is side-connected to the leads, 
the conduction electrons are back-scattered, resulting in zero conductance. \cite{note-inter}
Panel (c) shows that,  at the particle-hole symmetric point $V_g=-U/2$,  
the bonding orbital is doubly occupied, lying below the Fermi energy ${\rm E_{\rm F}}$, 
and the anti-bonding orbital, lying above ${\rm E_{\rm F}}$, is empty. Therefore, in this case, the 
double-dot structure creates no back-scattering density of states at ${\rm E_{\rm F}}$ [as schematically indicated 
in the panel (c)], resulting in perfect conductance. Finally, panel (d) displays the corresponding Kondo effect 
for the anti-bonding orbital, also resulting in zero conductance. 

\begin{figure}[t]
\includegraphics[width=3.4in]{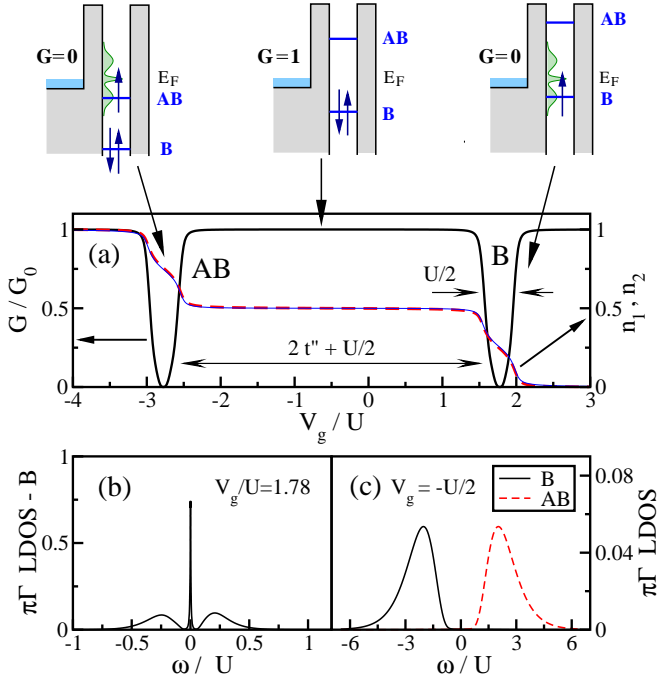}
\caption{(Color online) (a) Conductance and charge as a function of $V_{\rm g}$
for the molecular regime ($U=1.0$, $t^{\prime}=0.3$, 
and $t^{\prime \prime} = 2.0$) for the two side-connected quantum dots.
Notice that each feature of the conductance is associated to the respective diagram 
introduced in Fig.~\ref{cartoon-mol}. 
Panel (b) displays the LDOS for the bonding state (B) for $V_{\rm g}/U=1.78$ (the anti-bonding
state (AB) has an identical peak for $V_{\rm g}/U=-2.78$). Panel (c) displays the LDOS for 
the bonding (solid (black) line) and anti-bonding (dashed (red) line) states at the particle-hole symmetric 
point $V_g=-U/2$. 
$\lambda=2$ for all calculations.}
\label{fig:DOS-2qdA}
\end{figure}

Figure~\ref{fig:DOS-2qdA} shows the conductance and the  charge vs. gate voltage
in the molecular regime  as obtained with LDECA  for $t^{\prime \prime}=2.0$, $t^{\prime} = 0.3$,
and $U=1$ on a cluster with $L=11$ sites and $\lambda=2$. 
In this plot, one observes the two Fano-Kondo anti-resonances, with an approximate 
width of $U/2$, originating from two `molecular' levels separated by $\approx 2 \times t^{\prime \prime} + U/2$. 
It is important to emphasize that in the molecular regime the two dots behave as a unique entity, providing 
an extra lateral path for the electrons to traverse when visiting the Kondo peak derived 
from the molecular orbital. This gives rise to the Fano antiresonance in the conductance 
appearing in Fig.~\ref{fig:DOS-2qdA}(a). As expected, both dots are charged almost
simultaneously, as one can see in Fig.~\ref{fig:DOS-2qdA}(a), with a dashed line for dot 1 and a  solid thin line 
for dot 2. In the top of Fig.~\ref{fig:DOS-2qdA}, the corresponding potential wells described in 
Fig.~\ref{cartoon-mol} are displayed. 

To illustrate the idea of a `molecular orbital Kondo effect',
in the lower left panel, we display the LDOS associated with the molecular bonding orbital 
formed with the two dots. This LDOS is calculated at the positive gate potential at which  the conductance
is zero, which turns out to be at $V_g \approx t^{\prime \prime} - U/4$, as expected. We  find
the `molecular' Kondo peak at the Fermi energy, as well as the broadened $\epsilon$ 
and $\epsilon + \tilde{U}$ levels, where the renormalized intra-orbital 
Coulomb repulsion $\tilde{U}=U/2$ can be obtained by rewriting the dot Coulomb 
repulsion in the basis of the bonding and anti-bonding orbitals. A similar result (not shown) is found 
for the LDOS of the {\it anti}bonding orbital at the $V_g$ value for which 
the second Kondo-Fano resonance occurs. 
The LDOS for each quantum dot (not shown) also exhibits a Kondo peak.
Indeed, since the two dots equally participate in the molecular Kondo effect, their LDOS are quite similar
to each other, and qualitatively similar to what is shown in Fig.~\ref{fig:DOS-2qdA}(b).
Finally, in  Fig.~\ref{fig:DOS-2qdA}(c), we show 
the LDOS for both the bonding and anti-bonding orbitals for $V_g=-U/2$, where clearly 
the Kondo peak is absent and the two orbital levels are separated by about $2t^{\prime \prime}$. 

\begin{figure}
\vspace{3mm}
\includegraphics[width=3.4in]{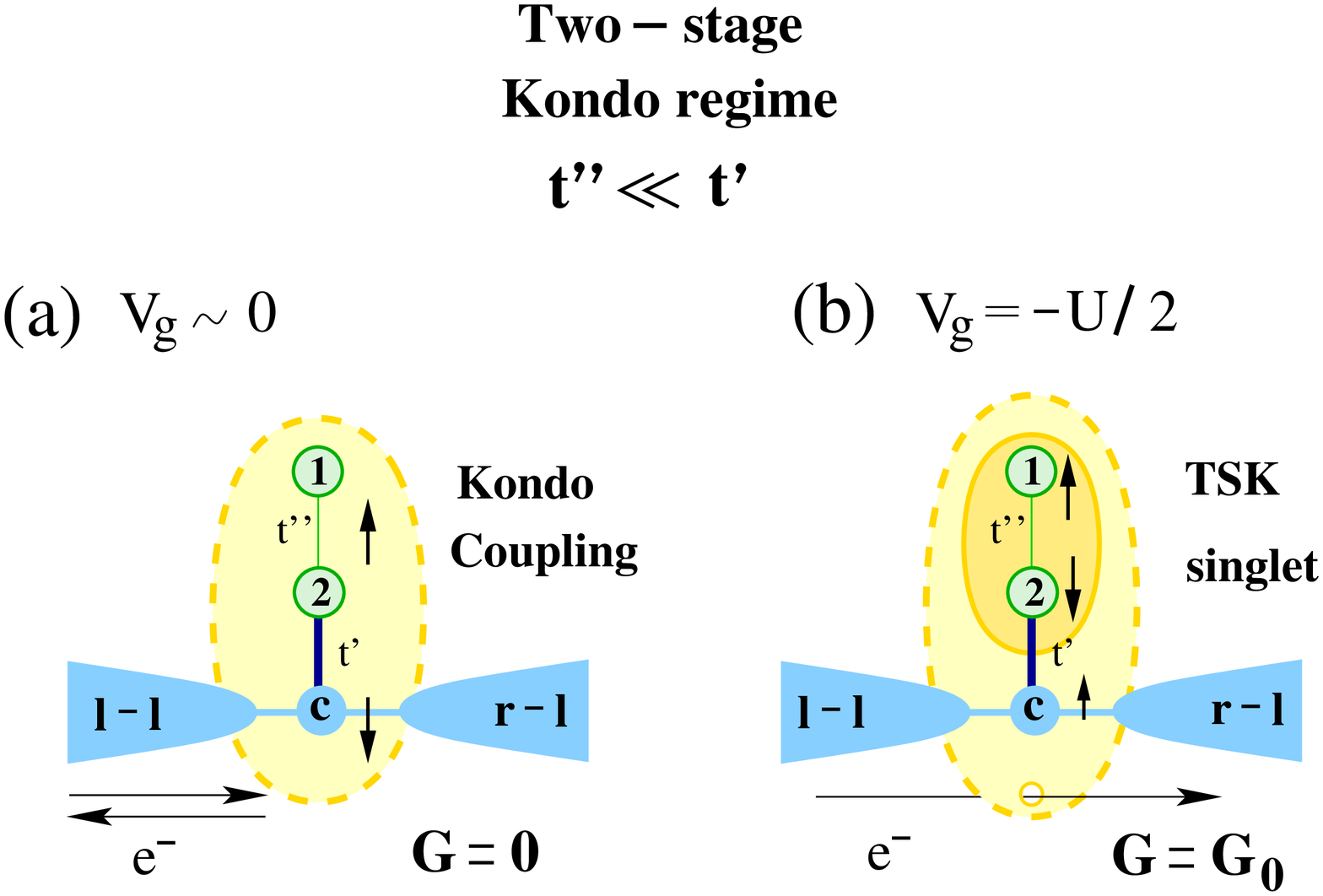}
\caption{(Color online) Schematic representation of the TSK regime, obtained
when $t^{\prime \prime} \ll t^{\prime}$. (a) Sketch of the first Kondo stage,
when the two dots are occupied by a single electron, which forms a Kondo singlet
with the conduction electrons, represented by the vertical antiparallel arrows 
(one in the dots, the other in the band). The Kondo
coupling, which mediates the formation of the many-body singlet, is represented
by the (yellow) lightly shaded oval. The density of states created by the Kondo effect
in dot 2 at the Fermi energy induces backward scattering of the conduction electrons
going through the leads (horizontal antiparallel arrows). This results in zero conductance. 
(b) At $V_g=-U/2$, when the dots have total charge $n=2$, 
the spins in the dots are locked into a singlet (darker shaded
oval). This singlet  is mediated through two Kondo effects, 
represented by the underlying lighter (yellow) oval. The combination of these two
Kondo effects, in contrast to the case in panel (a), removes density of states from
dot 2 exactly at the Fermi energy, suppressing the back-scattering and resulting in perfect
conductance. This is represented by the rightward arrow `piercing' the Kondo coupling oval.}
\label{cartoon-tsk}
\end{figure}

\begin{figure}
\vspace{3mm}
\includegraphics[width=3.4in]{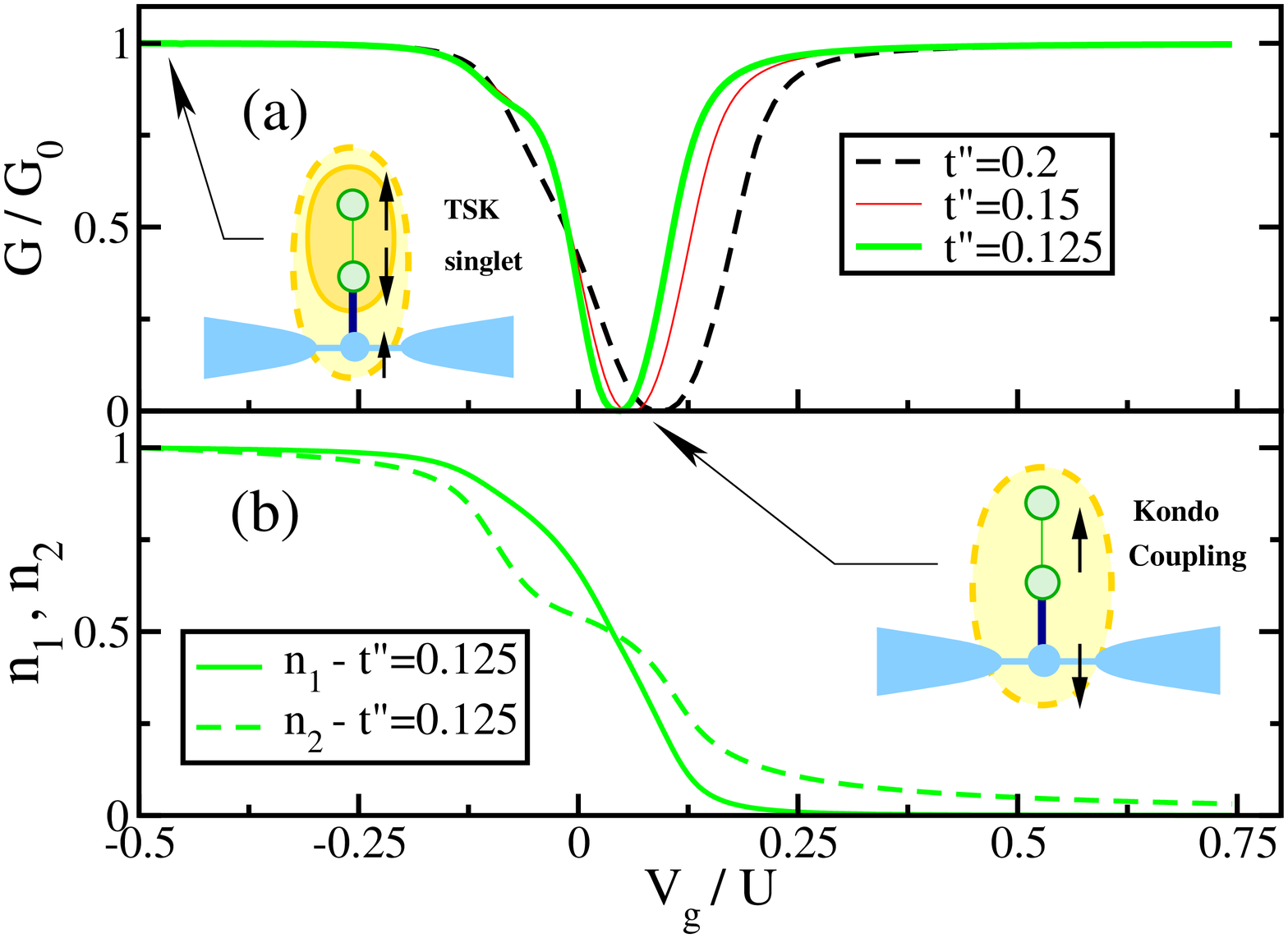}
\caption{(Color online) (a) Conductance and (b) charge in the dots as a function of $V_{\rm g}$
for values of $t^{\prime \prime} < t^{\prime}$ ($U=1.0$, $t^{\prime}=0.3$, and values of
$t^{\prime \prime}$ as indicated). See text for discussion. All calculations are for
$\lambda=2.0$, except for $t^{\prime \prime}=0.125$, where $\lambda=\sqrt{6}$.
Notice that the conductance and charge features at $V_g \approx 0$ and $V_g=-U/2$
were associated to the schematic diagrams introduced in Fig.~\ref{cartoon-tsk}.}
\label{fig:DOS-2qdB}
\end{figure}

\subsection{Two-stage Kondo regime}
\label{sec:tsk}

Figure~\ref{cartoon-tsk} schematically depicts a much more subtle regime 
than that of Fig.~\ref{cartoon-mol}, the TSK regime. One enters into this regime when 
$t^{\prime \prime} \ll t^{\prime}$, where now 
the connection of dot 2 with the leads is much stronger than the 
inter-dot connection. Here, the concept of bonding and anti-bonding 
orbitals does not apply, since each dot feels the interaction with 
the conduction electrons differently, the crucial point being that 
dot 1 interacts with the Fermi sea through dot 2.

In this and the next section, we will present 
evidence that our LDECA results are perfectly consistent with  the notion of TSK behavior. As a guidance to interpreting 
the numerical results, this behavior can be schematically described as follows. 
In Fig.~\ref{cartoon-tsk}(a), where the gate potential $V_g$ is such that 
the charge occupancy of the two dots is 1, {\it i.e.}, $n_1+n_2=1$, a Kondo effect develops, represented 
by the oval shape with dashed borders, resulting in back-scattering and a vanishing conductance, 
as indicated by the horizontal arrows. The Kondo effect involves a magnetic moment located 
in the dots, which is screened by the conduction electrons, indicated by the arrow on the dot and 
an antiparallel one in the band.
In Fig.~\ref{cartoon-tsk}(b), depicting the situation for $V_g=-U/2$, the two dots are each singly occupied 
and a strong singlet forms between them, represented by the darker oval shape with a solid border. 
Although in this regime the LDOS of dot 2 is exactly zero at the Fermi energy, therefore suppressing the back-scattering 
and restoring perfect conductance (this is represented by the arrow `piercing'
the lighter shaded oval), it does not eliminate the Kondo 
spin-spin correlation between the spin of dot 2 and the conduction spins. 
On the one hand, the conduction electrons do not see the two dot system 
as a unique entity: Indeed, they recognize dot 2 as a separate object to 
which their spins correlate. On the other hand, the spin of dot 1 sees the rest 
of the system as a whole, and Kondo correlates with the spin of dot 2, thus creating the two-dot singlet state. 
In reality, the singlet is a many-body effect {\it involving} 
the conduction electrons and it is composed of two consecutive Kondo effects (represented 
here by the underlying lightly shaded oval). 

In Fig.~\ref{fig:DOS-2qdB}, we show the LDECA conductance [panel (a)] and the charge in each dot [panel (b)] as a function of $V_g$ for
much lower values of $t^{\prime \prime}$ than in the previous molecular regime,
namely $t^{\prime \prime}=0.2$, $0.15$, and $0.125$. Let us first discuss the charge, as an example of a quantity
 that exhibits a qualitatively different behavior for high and low values of $t^{\prime \prime}/t^{\prime}$, 
with the two dots behaving more independently as $t^{\prime\prime} \to 0$.

In the small $t^{\prime\prime}/t^{\prime}$ regime, dot 2 is charged first upon approaching
$V_g=0$, and only when it has a substantial amount of charge
dot 1 starts to be charged as well [see Fig.~\ref{fig:DOS-2qdB}(b)]. In addition, around  $V_g\approx 0$
where the minimum in the conductance occurs due to the single-stage Kondo effect,  
the curve  for the charge of dot 2 features a much more well defined plateau than that for dot 1. 
This indicates that, in contrast to the molecular regime [see charge behavior in Fig.~\ref{fig:DOS-2qdA}(a)], 
the two dots now start to have a qualitatively different participation in the Kondo effect, 
suggesting that at this much lower value of $t^{\prime\prime}/t^{\prime}$ and at $V_g \approx 0$,
one starts to see the emergence of the first stage of the TSK regime. (Notice that 
DMRG results for the charge and the total spin as a function of $V_g$ 
for large clusters (not shown) agree with the LDECA picture just described).
In addition, the width of the Kondo anti-resonance seen in Fig.~\ref{fig:DOS-2qdB}(a) is now substantially
smaller than $U/2$, which is the typical value found in the molecular regime, see Fig.~\ref{fig:DOS-2qdA}(a), 
although much larger than the intrinsic width of the dots' resonance states. 
Finally, as one approaches $V_g=-U/2$, and each dot now has one electron, the second stage of the TSK is reached. In this regime, 
through the mediation of the conduction electrons, the 
interdot-singlet is formed and, although it shows a 
two-Kondo peak structure, the LDOS of dot 2 is zero at 
the Fermi energy [see Fig.~\ref{ldos-tsk}(b)]. As a consequence, the 
system exhibits perfect conductance [see Fig.~\ref{fig:DOS-2qdB}(a)].

\subsection{Conductance and LDOS at $V_g=-U/2$}

\begin{figure}\includegraphics[width=3.in]{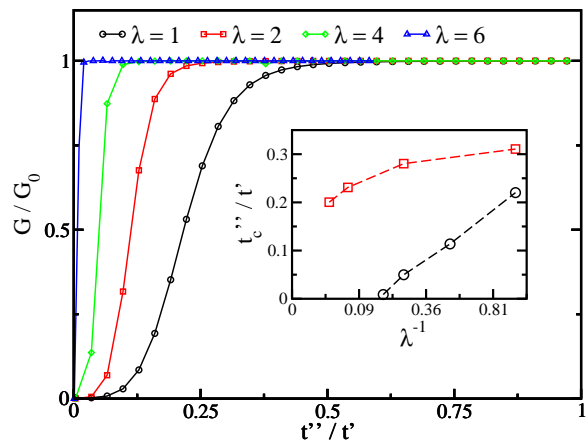}
\caption{(Color online) Conductance vs. $t^{\prime\prime}/t^{\prime}$ for $U=1.0$,
$t^{\prime}=0.3$, $V_g=-U/2$. Results are for $L=11$ and $\lambda=1$
 (black circles), $\lambda=2$ (red squares), $\lambda=4$ (green
diamonds) and $\lambda=6$ (blue triangles). In the inset, we
show $t^{\prime\prime}_c(\lambda)/t^{\prime}$ for $L=11$ (black circles)
and $L=5$ (red squares). 
}
\label{fig:2}
\end{figure}

Next, we discuss the conductance at the particle-hole symmetric
point as a function of $t^{\prime\prime}$ to  show that LDECA correctly captures the 
low-energy physics down to small values of $t''$.

Figure~\ref{fig:2} displays the conductance vs. $t^{\prime\prime}$
for various values of $\lambda$ and $L=11$.
We recall that, as exemplified in Figs.~\ref{fig:DOS-2qdA} and \ref{fig:DOS-2qdB},
the conductance at $V_g = -U/2$ for any $t^{\prime\prime}>0$ 
should be $G=G_0$. 
We study the electron-hole symmetry situation, as it is the 
most difficult point to be correctly described, having the 
lowest Kondo temperature for the set of parameters taken. 
The suppression of the conductance for small values of $t^{\prime\prime}/t^{\prime}$ shown in Fig.~\ref{fig:2}
is caused by finite-size effects which obscure the second stage Kondo effect. In this specific case, we find
the tendency of a strong suppression
of spin fluctuations in dot~1 as the system approaches half-filling, 
($V_g = -U/2$). This behavior at $\lambda=1$ is similar to other models discussed in detail
in Ref.~\onlinecite{hm08}. Figure~\ref{fig:2} suggests that the
finite-size dependence of the conductance for $\lambda=1$ (circles) is quite severe, as it starts to manifest 
itself at $t^{\prime\prime}/t^{\prime}\approx 0.5$.  However, it is also
evident that by increasing $\lambda$ the situation improves markedly, which is the main 
message to be taken from this figure. 

From the curves for each different $\lambda$ we can extract a characteristic inter-dot coupling
$t_c^{\prime \prime}(\lambda)$, satisfying $G(t_c^{\prime\prime})/G_0=1/2$, below which the 
conductance rapidly approaches zero.
The dependence of $t^{\prime\prime}_c$ on $\lambda^{-1}$ for two different values of $L$ is shown in the
inset. $t_c''(\lambda)$ tends to zero for values of $\lambda$ that decrease with increasing cluster size.
Therefore, at $V_{\rm g}$=$-U/2$, $G/G_0 \to 1$ when $t^{\prime\prime}\to 0$. 

\begin{figure}\includegraphics[width=3.in] {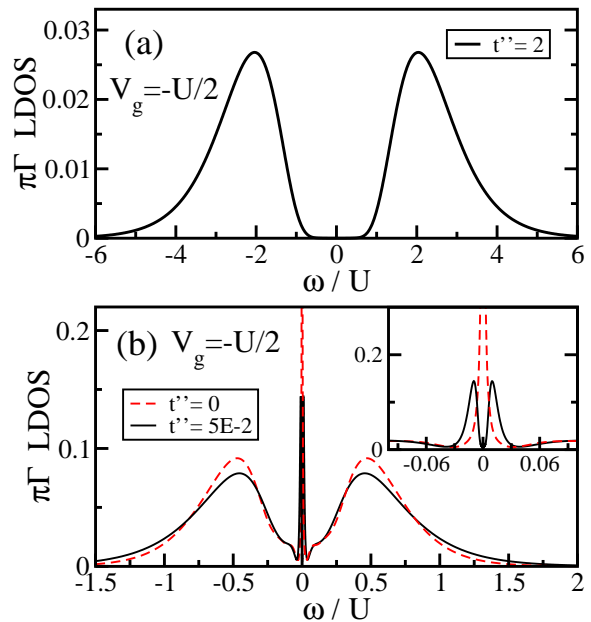}
\caption{(Color online) (a) LDOS at dot 2 for $L=11$, $t^{\prime\prime}=2$,
$t^{\prime}=0.3$, $U=1$, $V_g=-U/2$, and $\lambda=2$ (same parameters as the ones in
Fig.~\ref{fig:DOS-2qdA}(c), but here we show the LDOS of dot 2 only).
Notice the large region of vanishing LDOS around the Fermi energy ($\omega=0$).
(b) Same as in (a), but now for $t^{\prime\prime}=5 \times 10^{-2}$ and $\lambda=\sqrt{6}$ (solid (black) line).
Notice that, as in (a), the LDOS vanishes at the Fermi energy, but now only
in a very narrow interval. A comparison of the width of the double-peak structure
at low $\omega$ with the LDOS results for $t^{\prime\prime}=0$ (dashed (red) line),
{\it i.e.}, for a {\it single} side-connected dot, suggests that the double-peak structure is a 
split Kondo peak. An enlarged view of the double-peak structure is seen in the inset.
}
\label{ldos-tsk}
\end{figure}

 To further demonstrate the difference between the molecular and the TSK regimes at $V_g=-U/2$, Fig.~\ref{ldos-tsk} 
shows a comparison between the LDOS for two widely different 
values of $t^{\prime \prime}/t^{\prime}$. In Fig.~\ref{ldos-tsk}(a), we show the LDOS {\it at dot 2} 
for $t^{\prime\prime}=2.0$, $t^{\prime}=0.3$, $U=1.0$, and $\lambda=2$. These are the same parameters as 
the ones used in Fig.~\ref{fig:DOS-2qdA}(c), where the bonding and anti-bonding orbitals where shown. 
Figure \ref{ldos-tsk}(a)  unveils why the conductance in the molecular regime is $G=G_0$ at $V_g=-U/2$ 
[see Fig.~\ref{fig:DOS-2qdA}(a)], as dot 2 
has a vanishing density of states in a wide energy region around the Fermi level. 
Once there is no back-scattering density of states at the Fermi energy, the conductance 
is perfect. 

On the other hand, in  Fig.~\ref{ldos-tsk}(b), LDOS results for dot 2 are depicted  
for $t^{\prime\prime}=5 \time 10^{-2}$, and $\lambda=\sqrt{6}$, and the same values of $t^{\prime}$ and $U$ as in  Fig.~\ref{ldos-tsk}(a). 
Again, the density of states at the Fermi energy vanishes. However, in this case, close to the Fermi energy, 
we find two sharp features, suggestive of a Kondo peak split in two. To substantiate this picture, 
the dashed (red) curve in Fig.~\ref{ldos-tsk}(b) shows the Kondo peak that is present when $t^{\prime \prime}=0$, {\it i.e.}, 
when dot 1 is effectively removed. It is the presence of dot~1, interacting with the 
rest of the system through dot~2, that gives rise to the TSK regime, 
reflected in the LDOS of dot~2 as an antiresonance in the middle of its Kondo peak. 
 
In summary, the fact that the Kondo regime of dot 1 is mediated by the Kondo state of
dot 2 explains the  surprising result of a perfect conductance at $V_g=-U/2$ in the TSK regime.
This mechanism reduces the LDOS at the Fermi level of dot 2 to zero
as shown in Fig.~\ref{ldos-tsk}(b), eliminating an alternative path for 
the circulating electrons and hence any destructive interferences. In this
regime, the electrons at the dots form a spin singlet, even at small $t^{\prime\prime}/t^{\prime}$.
This subtle effect and its consequences on the conductance  are well captured by LDECA.
It is then clearly shown in Figs.~\ref{fig:2} and \ref{ldos-tsk} that 
the logarithmic discretization of the band, combined with the embedding process, 
provides reliable results in a wide parameter range.

\section{Summary}
\label{sec:sum}

In this paper, we developed a formalism to study local 
and highly correlated electrons that combines the numerical simplicity of
the ECA method with Wilson's idea of a logarithmic discretization of the non-interacting band.
A diagrammatic expansion that provides a solid theoretical basis for the method was also discussed.
Applied to a one-impurity problem, LDECA yields an excellent
agreement with BA results.
In addition, following the same procedure used in NRG
to broaden the LDOS, a perfect agreement was found
with the accepted results for the LDOS of the Anderson impurity model.
In the case of a double-dot side-connected to a lead, and at small $t^{\prime\prime}$, the contrast
between the $\lambda$=$1$ (ECA) and $\lambda>1$ (LDECA) results
exemplifies the power of the logarithmic discretization: the
low-energy physics associated to the TSK regime is
correctly unveiled, as LDECA provides an accurate description of the
physics close to the Fermi energy.

The main advantage of LDECA is its great flexibility, which allows
the incorporation of other degrees of freedom, such as localized
phonons or photons.\cite{Guille} The restrictions imposed by the
Lanczos method can be overcome by using DMRG,\cite{white} allowing one
to use larger systems and to study more involved problems. Related
efforts are in progress.

We conclude that the LDECA method can be applied to complex
problems, including  molecules adsorbed at a metallic surface
\cite{quique} and sophisticated topologies of quantum dots,
displaying exotic Kondo regimes, such as, for example, 
non-Fermi liquid behavior, two-channel Kondo effect and the physics associated to SU(N) systems.
\\

 {\bf Acknowledgments - } We thank K.A. Al-Hassanieh, L.G.G.V. Dias daSilva, E. Louis, V. Meden,
N. Sandler, A. Schiller, and J. A. Verges for helpful discussions.
E.V.A. and M.A.D. acknowledge the support of FAPERJ and CNPq (Brazil).
G.Ch. acknowledges financial support by the Spanish MCYT (grants
 FIS200402356, MAT2005-07369-C03-01
and NAN2004-09183-C10-08 and a Ram\'on y Cajal fellowship),
the Universidad de Alicante and the Generalitat Valenciana (grant
 GRUPOS03/092).  E.D. and F.H.-M. were supported
by the NSF grant DMR-0706020 and by the Division of Materials Science
 and Engineering, U.S. DOE under contract with UT-Battelle, LLC.
G.B.M. acknowledges support from NSF (DMR-0710529) and from
Research Corporation (Contract No. CC6542).

\appendix
\section{Diagrammatic Expansion}
\label{sec:app}

 In this appendix we develop the diagrammatic expansion of the one particle
 Green function at
 the impurity site given by
\begin{equation}
G_{00,\sigma}(t-t')=\frac{\left\langle
\mathcal{T}\{c_{0\sigma}(t)c^{\dagger}_{0\sigma}(t')S(\infty)\}\right\rangle_0}{\left\langle
S(\infty)\right\rangle_0}\,, \label{eq1}
\end{equation}
where, as usual, $S(\infty)$ is the evolution operator and $\mathcal{T}$ is
the time order operator.  The mean values are calculated in the
ground state of the unperturbed Hamiltonian $H_0$, given by Eq.~(\ref{Hzero}), restricting our
discussion to zero temperature.

The evolution operator $S(\infty)$ is expanded in increasing orders
of the perturbing term $H_{\mbox{\small p}}$, which, when inserted in Eq.~(\ref{eq1}),
gives rise to a perturbation series for the Green function.
It can be written as
\begin{widetext}
\begin{equation}
S(\infty)=\sum^\infty_{n=1}\frac{-i^n}{n!\big<S(\infty)\big>}\int^t_{t_0}\ldots\int^t_{t_0}
\left\langle \mathcal{T}\left\{c_{0\sigma}(t) H_{\mbox{\small p}}(t_1) \ldots
H_{\mbox{\small p}}(t_n)\right\}c^{\dagger}_{0\sigma}(t')\right\rangle_0
dt\ldots dt_n\, . \label{eq2}
\end{equation}

Substituting this expression into Eq.~(\ref{eq1}), the local Green
function is given by
\begin{eqnarray}
G_{00,\sigma}(t-t')&&=g_{0,\sigma}(t-t')+V^2\sum_{\sigma_1\sigma_2}\int
g_{0,N,N+1,\sigma,\sigma_1,\sigma_2}(t,t_1, t_2, t')dt_1
dt_2\nonumber\\
&&+V^4\sum_{\sigma_1\sigma_2\sigma_3\sigma_4}g_{0,N,N+1,\sigma,\sigma_1,\sigma_2,\sigma_3,\sigma_4}(t,t_1,
t_2, t_3, t_4,t')dt_1 dt_2 dt_3 dt_4+\nonumber\\
&&+\cdots \,.\label{eq3}
\end{eqnarray}
\end{widetext}
It is important to emphasize, as shown in Eq.~(\ref{eq3}), that
the conservation of charge of the unperturbed subsystem restricts the
expansion to even orders in $V$. The undressed Green function
appearing in the equation is defined as
\begin{equation}
\begin{array}{l}
g_{0,N,N+1,\sigma,\sigma_1,\sigma_2}(t,t_1,t_2,t')=\\ \\
\left\langle\!
\mathcal{T}\!\!\left\{\!c_{0\sigma}\!(t)c^{\dagger}_{N\sigma_1}\!(t_1\!)c_{N+1
\sigma_1}\!(t_1\!)c^{\dagger}_{N+1\sigma_2}\!(t_2\!)c_{N\sigma_2}\!(t_2\!)c^{\dagger}_{0\sigma}\!(t')\!\right\}\!\right\rangle_0
\end{array}
\label{eq4}
\end{equation}
with an obvious generalization for the undressed Green function of
other orders. Calculating terms of all orders in $V$ in the
expansion, Eq.~(\ref{eq3}), the local Green function can be
written as
\begin{equation}
G_{00,\sigma}(t-t')\!=\!G^{(0)}_{00,\sigma}(t-t')+G^{(2)}_{00,\sigma}(t-t')+G^{(4)}_{00,\sigma}(t-t')+\cdots.
\label{eq5}
\end{equation}
As the
operators belonging to the two different unperturbed parts of the
system are, in this ground
state, decoupled from each other since there is no connection between the cluster and the rest of
the leads, Eq.~(\ref{eq4}) results in
\begin{eqnarray}
&&g_{0,N,N+1,\sigma,\sigma_1,\sigma_2}(t,t_1,t_2,t')\nonumber \\
&&=g_{0,N,\sigma,\sigma_1}(t,t_1,t_2,t')g_{N+1,\sigma_1}
(t_1,t_2)\delta_{\sigma_1\sigma_2} \label{eq6}
\end{eqnarray}
where the spin conservation imposes the condition
$\sigma_1=\sigma_2$ and
\begin{subequations}
\begin{eqnarray}
g_{0,N,\sigma,\sigma_1}\!(t,t_1,t_2,t')\!\!\!&=&\!\!\!\left\langle
\!\mathcal{T}\{c_{0,\sigma}(t)c^{\dagger}_{N,\sigma_1}\!(t_1\!)c_{N,
\sigma_1}\!(t_2)c^{\dagger}_{0,\sigma}(t')\!\}\!\!\right\rangle_{\!\!0}
\nonumber\\ &&\label{eq7a}\\
g_{N+1,\sigma_1}\!(t_1,t_2)\!\!\!&=&\!\!\!\left\langle
\mathcal{T}\{c_{N+1,\sigma_1}\!(t_1\!)c^{\dagger}_{N+1,\sigma_1}\!(t_2)\}\right\rangle_0\,.
\label{eq7b} \label{eq7}
\end{eqnarray}
\end{subequations}
According to Eq.~(\ref{eq6}), the expansion Eq.~(\ref{eq3}) is formally
a locator-propagator expansion \cite{metzner} where the
locators correspond to the unperturbed sub-systems Green functions
and the propagator turns out to be the one connecting them.

The Green function $g_{0,N,\sigma,\sigma_1}$ can be diagrammatically
represented by
\begin{eqnarray}
\nonumber \\
g_{0,N,\sigma,\sigma_1}(t,t_1,t_2,t')&=&\hskip2.5cm
\begin{picture}(0,0)(75,0)
\put(8,3){\line(1,0){10}} \put(8,3){\line(1,0){10}}
\put(5,3){\circle{6}} \put(16,13){$\sigma_1$}
\put(18,1){\framebox(39,5)} \put(57,3){\line(1,0){10}}
\put(70,3){\circle{6}} \put(18,25){$t_1$} \put(29,25){\circle*{6}}
\put(29,6){\line(0,1){16}} \put(53,25){$t_2$}
\put(49,25){\circle*{6}} \put(49,6){\line(0,1){16}}
\put(52,13){$\sigma_1$} \put(1,-10){$t$} \put(36,-10){$\sigma$}
\put(67,-10){$t'$}
\end{picture}
\nonumber \\ &&
\nonumber \\ &&
\mathcal{T}\left\{c_{o\sigma}(t)c^{\dagger}_{N\sigma_1}(t_1)c_{N\sigma_1}(t_2)c^{\dagger}_{o\sigma}(t')\right\}
\nonumber
\end{eqnarray}
and
\begin{equation}
g_{N+1,\sigma}(t_1,t_2)=\hskip1.3cm
\begin{picture}(0,0)(40,0)
\put(7,2){\circle*{6}} \put(7,2){\begin{xy}<1.3cm,0cm> \ar@{~}(2,0)
\end{xy}}
\put(36,2){\circle*{6}} \put(3,10){$t_1$} \put(18,10){$\sigma$}
\put(34,10){$t_2$}
\end{picture}
\mathcal{T}\left\{c_{N+1\sigma_1}(t_1)c_{N+1\sigma_1}(t_2)\right\}\nonumber
\label{e7}
\end{equation}
The zeroth-order contribution to the Green function, {\it i.e.}, the solution of
the problem for $V=0$, is represented in terms of diagrams as
\begin{equation}
G_{00,\sigma}^{(0)}(t-t')=\left\langle
\mathcal{T}\left\{c_{0\sigma}(t)c^{\dagger}_{0\sigma}(t')\right\}\right\rangle_0 =
\put(8,3){\line(1,0){10}} \put(5,3){\circle{6}}
\put(18,1){\framebox(26,5)} \put(44,3){\line(1,0){10}}
\put(57,3){\circle{6}} \put(3,-9){$t$} \put(26,-9){$\sigma$}
\put(52,-9){$t'$} \qquad\qquad\qquad \label{eq8}
\end{equation}

The Green function is exactly obtained by calculating the
ground state of the cluster using the Lanczos method. Although it
includes the many-body interaction and its effects within the
cluster, it is the undressed Green function with respect to the expansion
given by Eq.~(\ref{eq5}).

To second order in perturbation theory, the contribution to
$G^{}_{00,\sigma}(t-t')$ is
\begin{eqnarray}
&&G^{(2)}_{00,\sigma}(t-t')=\frac{V^2}{\left<
S(\infty)\right>_{0}}\nonumber\\
&&\sum_{\sigma_1}\!\!\int\!\!
g_{0,N,\sigma_,\sigma_1}(t,t_1,t_2,t')g_{N+1,\sigma_1}(t_1,t_2)dt_1dt_2
\label{eq9}
\end{eqnarray}
and can be diagrammatically represented as
\begin{equation}
G^{(2)}_{00,\sigma}(t-t')=\hskip2.8cm
\begin{picture}(0,30)(80,0)
\put(8,3){\line(1,0){10}} \put(5,3){\circle{6}}
\put(14,13){$\sigma_1$} \put(2,-10){$t$} \put(13,25){$t_1$}
\put(30,32){$\sigma_1$} \put(26,25){\circle*{6}}
\put(26,6){\line(0,1){16}} \put(26,25){\begin{xy}<1cm,0cm>
\ar@{~}(2,0)
\end{xy}}
\put(50,25){$t_2$} \put(30,-9){$\sigma$} \put(46,25){\circle*{6}}
\put(46,6){\line(0,1){16}} \put(49,13){$\sigma_1$}
\put(18,1){\framebox(37,5)} \put(55,3){\line(1,0){10}}
\put(68,3){\circle{6}} \put(72,-10){$t'$}
\end{picture}\,,
\label{eq10}
\end{equation}
where an integration over each internal time $t_1$ and $t_2$ is implied
and the sum over
$\sigma_1$ needs to be taken.

Regarding the calculation of the many-particle Green functions at
the lead sites outside the cluster, Wick's theorem can be used
since this part of the system is represented by a one-body
Hamiltonian. In this case, it is clear that for the fourth order
we obtain the diagram
\begin{equation}
G^{(4)}_{00,\sigma}(t-t')=\hskip4.2cm
\begin{picture}(0,25)(120,0)
\put(8,3){\line(1,0){10}} \put(5,3){\circle{6}}
\put(14,13){$\sigma_1$} \put(2,-10){$t$} \put(13,25){$t_1$}
\put(30,32){$\sigma_1$} \put(26,25){\circle*{6}}
\put(26,6){\line(0,1){16}} \put(26,25){\begin{xy}<1cm,0cm>
\ar@{~}(2,0)
\end{xy}}
\put(50,25){$t_2$} \put(60,-9){$\sigma$} \put(46,25){\circle*{6}}
\put(46,6){\line(0,1){16}} \put(49,13){$\sigma_1$}
\put(18,1){\framebox(90,5)} \put(108,3){\line(1,0){10}}
\put(121,3){\circle{6}} \put(120,-10){$t'$} \put(65,13){$\sigma_2$}
\put(65,25){$t_3$} \put(78,25){\circle*{6}}
\put(78,6){\line(0,1){16}} \put(78,25){\begin{xy}<1cm,0cm>
\ar@{~}(2,0)
\end{xy}}
\put(80,32){$\sigma_2$} \put(104,25){$t_4$} \put(98,25){\circle*{6}}
\put(98,6){\line(0,1){16}} \put(101,13){$\sigma_2$}
\end{picture}\,,
\label{eq11}
\end{equation}
such that the dressed locator $G_{00,\sigma}(t-t')$ can be cast into
\begin{widetext}
\bigskip
\begin{equation}
G_{00,\sigma}(t-t')=\hskip14cm
\begin{picture}(0,0)(400,0)
\put(8,3){\line(1,0){10}} \put(5,3){\circle{6}}
\put(18,1){\rule{1cm}{2mm}} \put(45,3){\line(1,0){10}}
\put(58,3){\circle{6}} \put(3,-10){$t$} \put(29,-10){$\sigma$}
\put(56,-10){$t'$} \put(65,0){$=$}
\end{picture}
\begin{picture}(0,0)(325,0)
\put(8,3){\line(1,0){8.2}} \put(5,3){\circle{6}}
\put(17,1){\framebox(27,6)} \put(44,3){\line(1,0){8.2}}
\put(56,3){\circle{6}} \put(3,-10){$t$} \put(29,-10){$\sigma$}
\put(54,-10){$t'$} \put(65,0){$+$}
\end{picture}
\begin{picture}(0,0)(250,0)
\put(8,3){\line(1,0){10}} \put(5,3){\circle{6}}
\put(18,1){\framebox(42,5)} \put(60,3){\line(1,0){10}}
\put(73,3){\circle{6}} \put(29,25){\circle*{6}}
\put(29,6){\line(0,1){16}} \put(29,25){\begin{xy}<1cm,0cm>
\ar@{~}(2,0)
\end{xy}}
\put(49,25){\circle*{6}} \put(49,6){\line(0,1){16}} \put(3,-10){$t$}
\put(37,-10){$\sigma$} \put(69,-10){$t'$} \put(14,25){$t_1$}
\put(14,13){$\sigma_1$} \put(53,25){$t_2$} \put(50,13){$\sigma_1$}
\put(31,30){$\sigma_1$} \put(80,0){$+$}
\end{picture}
\begin{picture}(0,0)(160,0)
\put(9,3){\line(1,0){10}} \put(6,3){\circle{6}}
\put(19,1){\framebox(85,5)} \put(104,3){\line(1,0){10}}
\put(117,3){\circle{6}} \put(27,25){\circle*{6}}
\put(27,6){\line(0,1){16}} \put(26,25){\begin{xy}<1cm,0cm>
\ar@{~}(2,0)
\end{xy}}
\put(47,25){\circle*{6}} \put(47,6){\line(0,1){16}}
\put(78,25){\circle*{6}} \put(78,6){\line(0,1){16}}
\put(78,25){\begin{xy}<1cm,0cm> \ar@{~}(2,0)
\end{xy}}
\put(97,25){\circle*{6}} \put(97,6){\line(0,1){16}} \put(3,-10){$t$}
\put(58,-10){$\sigma$} \put(113,-10){$t'$} \put(14,25){$t_1$}
\put(14,13){$\sigma_1$} \put(51,25){$t_2$} \put(49,13){$\sigma_1$}
\put(65,13){$\sigma_2$} \put(31,30){$\sigma_1$}
\put(82,30){$\sigma_2$} \put(65,25){$t_3$} \put(102,25){$t_4$}
\put(101,13){$\sigma_2$} \put(125,0){$+\cdots$ \ .}
\end{picture}
\label{eq12}
\end{equation}
\bigskip
\end{widetext}
Although the cluster's undressed one-particle Green function
$\begin{picture}(0,0)(3,0) \put(8,4){\line(1,0){10}}
\put(5,4){\circle{6}} \put(18,2){\framebox(26,5)}
\put(44,4){\line(1,0){10}} \put(57,4){\circle{6}} \put(3,-9){$^t$}
\put(29,-7){$^\sigma$} \put(55,-9){$^{t'}$}
\end{picture}\hskip2cm$
can be calculated exactly using the Lanczos method, the undressed cluster, $n$-particle Green function,
\[\begin{picture}(0,0)(40,10)
\put(8,3){\line(1,0){10}} \put(5,3){\circle{6}} \put(28,18){$^1$}
\put(18,1){\framebox(90,5)} \put(108,3){\line(1,0){10}}
\put(122,3){\circle{6}} \put(24,19){\circle*{6}}
\put(24,6){\line(0,1){10}} \put(55,18){$^2$}
\put(38,19){\circle*{6}} \put(38,6){\line(0,1){10}}
\put(89,18){$^n$} \put(51,19){\circle*{6}}
\put(51,6){\line(0,1){10}} \put(65,19){\circle*{6}}
\put(65,6){\line(0,1){10}} \put(68,9){$\cdots$}
\put(85,19){\circle*{6}} \put(85,6){\line(0,1){10}}
\put(99,19){\circle*{6}} \put(99,6){\line(0,1){10}} \put(130,2){, is
unknown for $n\geq 1$.}
\end{picture}\hskip5cm\]
It is clear that these functions cannot be calculated directly
using the prescription provided by  Wick's theorem because they
include many-body Coulomb contributions coming from the impurity.

In order to sort out this difficulty we propose another perturbation
expansion, assuming the cluster without the many-body term at the
impurity as the unperturbed Hamiltonian and $H_{MB}$ as the
perturbation. The enormous advantage of this expansion in contrast
with the previous one is that  Wick's theorem is applicable
because the non-perturbed system is represented by a one-particle
Hamiltonian. For an infinite system, this expansion has been
extensively used to calculate the one-particle Green function to
study, for instance, the Kondo effect. In most cases, these studies
have been restricted to expansions in the self-energy up to second
order in the Coulomb interaction parameter $U$.\cite{metzner} However,
we are in a different situation here because the system is finite and,
more importantly, it requires the calculation of the Green function to
all orders in the number of particles. In our case, the one particle
Green function can be numerically calculated. After these diagrams
are obtained, they are incorporated into the original diagrammatic
expansion, Eq.~(\ref{eq12}), in order to obtain the Green
function of the complete system $G_{00,\sigma}(t-t')$. When
calculating the self-energy, this procedure in principle permits  to
sum up, to all orders in $U$, the most important families of diagrams.
These are chosen among the ones that are essential to give a proper
account of the region near the Fermi level.

We use Eqs.~(\ref{eq1}) and (\ref{eq2}) to obtain this new
diagrammatic expansion. It is worth mentioning that now the mean
values $\langle \cdots \rangle_0$ are calculated in the ground state of the
cluster without the Coulomb interaction and that the evolution
operator Eq.~(\ref{eq2}) requires the substitution of $H_{p}$ by $H_{MB}$.

In order to clarify the procedure and to establish the diagrammatic
rules, we calculate the first diagrams corresponding to the locator
$g_{0,N,\sigma,\sigma_1}(t,t_1,t_2,t')$, Eq.~(\ref{eq7a}). We
define three undressed Green functions,
\begin{subequations}
\begin{equation}
g^{(0)}_{00,\sigma}(t-t')=\left\langle
\mathcal{T}\left\{c_{0\sigma}(t),c_{0\sigma}(t')\right\}\right\rangle_0
=\hskip2cm
\begin{picture}(0,0)(60,0)
\put(10,2){\circle{6}} \put(56,2){\circle{6}}
\put(13,2){\line(1,0){40}} \put(10,8){$t$} \put(54,8){$t'$}
\put(28,7){$\sigma$}
\end{picture}
\label{eq13a}
\end{equation}
\begin{equation}
g^{(0)}_{0N,\sigma}(t-t')=\left\langle
\mathcal{T}\left\{c_{0\sigma}(t),c^{\dagger}_{N\sigma}(t')\right\}\right\rangle_0
=\hskip1.9cm
\begin{picture}(0,0)(60,0)
\put(10,2){\circle{6}} \put(56,2){\circle*{6}}
\put(13,2){\line(1,0){40}} \put(10,8){$t$} \put(54,8){$t'$}
\put(28,7){$\sigma$}
\end{picture}
\label{eq13b}
\end{equation}
\begin{equation}
g^{(0)}_{N\sigma}(t-t')=\left\langle
\mathcal{T}\left\{c_{N\sigma}(t)c^{\dagger}_{N\sigma}(t')\right\}\right\rangle_0
=\hskip1.9cm
\begin{picture}(0,0)(60,0)
\put(10,2){\circle*{6}} \put(56,2){\circle*{6}}
\put(13,2){\line(1,0){40}} \put(10,8){$t$} \put(54,8){$t'$}
\put(28,7){$\sigma$}
\end{picture}
\label{eq13c}
\end{equation}
\label{eq13}
\end{subequations}
that, together with Eq.~(\ref{e7}), constitute the building
blocks of the diagrammatic expansion.

The contribution to the Green function to zero order in $U$,
$g_{0,N,\sigma,\sigma_1}(t,t_1,t_2,t')$, defined in Eq.~(\ref{eq7a}) is given by
\begin{equation}
g^{(0)}_{0N,\sigma}(t,t_1)g^{(0)}_{0N,\sigma}(t_2,t')\delta_{\sigma\sigma'}+
g^{(0)}_{00,\sigma}(t,t')g_{NN,\sigma'}(t_1,t_2)= \label{eq14}
\end{equation}
\begin{picture}(0,50)(-40,-10)
\put(13,-1){\circle{6}} \put(13,35){\circle*{6}}
\put(13,2){\line(0,1){30}} \put(2,-4){$t$} \put(1,30){$t_1$}
\put(47,-4){$t'$} \put(3,12){$\sigma$} \put(43,-1){\circle{6}}
\put(43,35){\circle*{6}} \put(43,2){\line(0,1){30}}
\put(47,30){$t_2$} \put(47,12){$\sigma$} \put(60,12){+}
\put(80,-2){\circle{6}} \put(121,-2){\circle{6}}
\put(83,-2){\line(1,0){35}} \put(80,4){$t$} \put(121,4){$t'$}
\put(98,3){$\sigma$} \put(80,32){\circle*{6}}
\put(121,32){\circle*{6}} \put(83,32){\line(1,0){35}}
\put(80,40){$t_1$} \put(121,40){$t_2$} \put(98,35){$\sigma'$}
\put(129,-2){.}
\end{picture}

From this result we infer that the contribution to
$G^{(2)}_{00,\sigma}(t-t')$ in zero order in $U$ is
\[
\begin{picture}(0,45)(80,0)
\put(13,-1){\circle{6}} \put(13,30){\circle*{6}}
\put(13,2){\line(0,1){30}} \put(2,-4){$t$} \put(1,30){$t_1$}
\put(47,-4){$t'$} \put(3,12){$\sigma$} \put(43,-1){\circle{6}}
\put(43,30){\circle*{6}} \put(43,2){\line(0,1){30}}
\put(47,30){$t_2$} \put(47,12){$\sigma$}
\put(13,30){\begin{xy}<1.4cm,0cm> \ar@{~}(2,0)
\end{xy}}
\put(27,37){$\sigma$} \put(62,12){+} \put(80,-2){\circle{6}}
\put(126,-2){\circle{6}} \put(83,-2){\line(1,0){40}} \put(80,4){$t$}
\put(124,4){$t'$} \put(80,32){\circle*{6}}
\put(78,33){\begin{xy}<1.8cm,0cm> \ar@{~}(2,0)
\end{xy}}
\put(125,32){\circle*{6}} \put(103,32){\oval(41,14)[b]}
\put(80,40){$t_1$} \put(124,40){$t_2$} \put(100,38){$\sigma'$}
\put(100,14){$\sigma'$} \put(100,0){$\sigma$} \put(134,-2){.}
\end{picture}
\]

The second diagram is a non-connected one and, as usual, does not
contribute to the Green function.\cite{Abrikosov}

To first order in $U$, incorporating all the possible contractions
resulting from the application of Wick's theorem and eliminating the
non-connected diagrams, the contributions to
$G^{(2)}_{00,\sigma}(t-t')$ are

\begin{equation}
\begin{picture}(0,50)(150,0)
\put(93,45){\begin{xy}<1.4cm,0cm>
\ar@{~}(2,0)
\end{xy}}
\put(84,-6){$t$}
\put(83,50){$t_1$}
\put(129,-6){$t'$}
\put(126,8){$\sigma$}
\put(90,8){$\sigma$}
\put(126,31){$\bar\sigma$}
\put(90,31){$\bar\sigma$}
\put(105,50){$\bar\sigma$}
\put(127,50){$t_2$}
\put(95,45){\circle*{6}}
\put(125,45){\circle*{6}}
\put(95,-1){\circle{6}}
\put(125,-1){\circle{6}}
\put(124,2){\line(-2,3){12}}
\put(108,20){\line(-2,-3){12}}
\put(108,25){\line(-2,3){12}}
\put(125,43){\line(-2,-3){12}}
\put(110,23){\circle{6}}
\put(118,23){$t_3$}
\put(145,22){+}
\put(173,-1){\circle{6}}
\put(173,45){\circle*{6}}
\put(173,2){\line(0,1){45}}
\put(162,-6){$t$}
\put(161,50){$t_1$}
\put(207,-6){$t'$}
\put(213,-6){$.$}
\put(163,27){$\sigma$}
\put(203,-1){\circle{6}}
\put(203,45){\circle*{6}}
\put(203,2){\line(0,1){45}}
\put(207,50){$t_2$}
\put(205,7){$\sigma$}
\put(170,45){\begin{xy}<1.4cm,0cm>
\ar@{~}(2,0)
\end{xy}}
\put(187,50){$\sigma$}
\put(203,25){\line(0,1){20}}
\put(203,2){\line(0,1){17}}
\put(210,22){\oval(12,10)[t]}
\put(210,22){\oval(12,10)[b]}
\put(204,22){\circle{6}}
\put(204,22){\color{white}\circle*{4.8}}
\put(219,20){$\bar\sigma$}
\put(205,30){$\sigma$}
\put(192,20){$t_3$}
\end{picture}\nonumber
\end{equation}
From these calculations,  we conclude that there are two different types 
of vertices $\ \put(4,2){\circle*{6}}\quad$ and $\
\put(4,2){\circle{6}}\quad$. At each $\ \put(4,2){\circle*{6}}\quad$
vertex there is one incoming and outgoing propagator and a 
factor of $V$ has to be included. These are the vertices that result from
the one particle Hamiltonian $H_{p}$. The other vertex $\
\put(4,2){\circle{6}}\quad$ comes from the Hamiltonian $H_{MB}$.
There are two incoming and two outgoing spin $\sigma$ and
$\bar\sigma$  propagators and a factor of $U$ included at this vertex.
As usual, the integral over the time variable associated to each vertex has to be
taken.

These rules are schematically  represented as
\[
\begin{picture}(0,50)(100,0)
\put(24,36){\begin{xy}<1.4cm,0cm> \ar@{~}(2,0)
\end{xy}}
\put(37,42){$\sigma$} \put(17,40){$V$} \put(30,25){$t_1$}
\put(15,16){$\sigma$} \put(26,1){\line(0,1){32}}
\put(27,35){\circle*{6}} \put(174,3){\line(-2,3){12}}
\put(158,20){\line(-2,-3){12}} \put(158,25){\line(-2,3){12}}
\put(175,43){\line(-2,-3){12}} \put(160,23){\circle{6}}
\put(175,33){$\bar\sigma$} \put(175,7){$\sigma$}
\put(142,7){$\bar\sigma$} \put(142,33){$\sigma$} \put(146,18){$t_1$}
\put(170,18){$U$} \put(180,18){$.$}
\end{picture}
\]
To second order in $U$ the topologically different connected
diagrams that contribute to $G_{00,\sigma}(t-t')$ are
\begin{equation}
\begin{picture}(0,0)(150,-90)
\put(203,-3){\circle{6}}
\put(203,45){\circle*{6}}
\put(203,0){\line(0,1){10}}
\put(203,35){\line(0,1){9}}
\put(203,16){\line(0,1){13}}
\put(192,-6){$t$}
\put(191,50){$t_1$}
\put(237,-6){$t'$}
\put(233,-3){\circle{6}}
\put(233,45){\circle*{6}}
\put(233,0){\line(0,1){45}}
\put(211,13){\oval(12,10)[b]}
\put(211,13){\oval(12,10)[t]}
\put(203,13){\circle{6}}
\put(203,13){\color{white}\circle*{4.8}}
\put(237,50){$t_2$}
\put(203,45){\begin{xy}<1.4cm,0cm>
\ar@{~}(2,0)
\end{xy}}
\put(195,32){\oval(12,10)[b]}
\put(195,32){\oval(12,10)[t]}
\put(203,32){\circle{6}}
\put(203,32){\color{white}\circle*{4.8}}
\put(204,2){$\sigma$}
\put(204,20){$\sigma$}
\put(204,36){$\sigma$}
\put(217,50){$\sigma$}
\put(237,20){$\sigma$}
\put(220,11){$\bar\sigma$}
\put(179,29){$\bar\sigma$}
\put(190,10){$t_3$}
\put(209,28){$t_4$}
\put(260,22){}
\end{picture}
\begin{picture}(0,0)(115,-90)
\put(13,-1){\circle{6}}
\put(13,45){\circle*{6}}
\put(13,2){\line(0,1){17}}
\put(13,25){\line(0,1){20}}
\put(2,-6){$t$}
\put(1,50){$t_1$}
\put(47,-6){$t'$}
\put(3,5){$\sigma$}
\put(-10,20){$\bar\sigma$}
\put(43,-1){\circle{6}}
\put(43,45){\circle*{6}}
\put(43,25){\line(0,1){20}}
\put(43,2){\line(0,1){17}}
\put(50,22){\oval(12,10)[t]}
\put(50,22){\oval(12,10)[b]}
\put(43,22){\circle{6}}
\put(43,22){\color{white}\circle*{4.8}}
\put(47,50){$t_2$}
\put(47,5){$\sigma$}
\put(13,45){\begin{xy}<1.4cm,0cm>
\ar@{~}(2,0)
\end{xy}}
\put(27,52){$\sigma$}
\put(7,22){\oval(12,10)[t]}
\put(7,22){\oval(12,10)[b]}
\put(13,22){\circle{6}}
\put(13,22){\color{white}\circle*{4.8}}
\put(3,32){$\sigma$}
\put(47,32){$\sigma$}
\put(60,20){$\bar\sigma$}
\put(16,15){$t_3$}
\put(32,15){$t_4$}
\put(70,22){}
\end{picture}
\begin{picture}(0,0)(125,-90)
\put(103,-1){\circle{6}}
\put(103,45){\circle*{6}}
\put(103,2){\line(0,1){8}}
\put(103,35){\line(0,1){9}}
\put(103,16){\line(0,1){13}}
\put(92,-6){$t$}
\put(91,50){$t_1$}
\put(137,-6){$t'$}
\put(93,5){$\sigma$}
\put(133,-1){\circle{6}}
\put(133,45){\circle*{6}}
\put(133,2){\line(0,1){45}}
\put(106,22){\oval(13,17)[r]}
\put(103,13){\circle{6}}
\put(137,50){$t_2$}
\put(103,45){\begin{xy}<1.4cm,0cm>
\ar@{~}(2,0)
\end{xy}}
\put(117,52){$\sigma$}
\put(100,22){\oval(13,17)[l]}
\put(103,32){\circle{6}}
\put(93,32){$\sigma$}
\put(137,20){$\sigma$}
\put(116,20){$\bar\sigma$}
\put(103,20){$\sigma$}
\put(85,20){$\bar\sigma$}
\put(106,5){$t_3$}
\put(106,32){$t_4$}
\put(160,22){}
\end{picture}
\begin{picture}(0,140)(260,0)
\put(195,45){\begin{xy}<1.4cm,0cm>
\ar@{~}(2,0)
\end{xy}}
\put(182,-6){$t$}
\put(181,50){$t_1$}
\put(232,-6){$t'$}
\put(220,9){$\sigma$}
\put(180,8){$\bar\sigma$}
\put(220,31){$\sigma$}
\put(195,31){$\sigma$}
\put(201,52){$\sigma$}
\put(227,50){$t_2$}
\put(195,45){\circle*{6}}
\put(224,45){\circle*{6}}
\put(195,-1){\circle{6}}
\put(226,0){\circle{6}}
\put(224,2.5){\line(-2,3){12}}
\put(208,25){\line(-2,3){12}}
\put(224,43){\line(-2,-3){12}}
\put(209,20){\line(-2,-3){12}}
\put(210,23){\circle{6}}
\put(205,5){$t_3$}
\put(207,14){$\sigma$}
\put(198,1){$\sigma$}
\put(218,23){$t_4$}
\put(245,22){}
\put(195,12){\oval(12,10)[t]}
\put(195,12){\oval(12,10)[b]}
\put(202,12){\color{white}\circle*{4.8}}
\put(202,12){\circle{6}}
\put(160,22){}
\put(280,45){\begin{xy}<1.5cm,0cm>
\ar@{~}(2,0)
\end{xy}}
\put(268,-7){$t$}
\put(268,53){$t_1$}
\put(325,-7){$t'$}
\put(318,6){$\sigma$}
\put(272,6){$\sigma$}
\put(314,33){$\sigma$}
\put(272,33){$\sigma$}
\put(295,52){$\sigma$}
\put(295,25){$\bar\sigma$}
\put(295,5){$\bar\sigma$}
\put(317,53){$t_2$}
\put(280,45){\circle*{6}}
\put(315,45){\circle*{6}}
\put(278,-2){\circle{6}}
\put(317,-2){\circle{6}}
\put(313,22){$t_4$}
\put(274,22){$t_3$}
\put(285,26){\line(-1,3){7}}
\put(285,22){\line(-1,-3){7}}
\put(317,1){\line(-1,3){7}}
\put(317,47){\line(-1,-3){7}}
\put(287,24){\circle{6}}
\put(297.5,21){\oval(22,13)[b]}
\put(297.5,27){\oval(22,13)[t]}
\put(308,24){\circle{6}}
\put(335,-7){.}
\end{picture}
\label{eq15}
\end{equation}

\vspace{3mm}
The  one-particle cluster Green function exactly obtained by
numerical means, defined in Eq.~(\ref{eq8}), can be thought of to
be the result of the sum of the following infinite series of diagrams:
\begin{eqnarray}
&&G^{(0)}_{i0,\sigma}(t-t')=\hskip6cm
\begin{picture}(0,0)(170,0)
\put(17,1){\framebox(30,5)} \put(7,3){\line(1,0){10}}
\put(4,3){\circle{6}} \put(1,1.5){\tiny$\times$} \put(1,-8){$t$}
\put(60,-8){$t'$} \put(2,8){$i$} \put(60,3){\circle{6}}
\put(47,3){\line(1,0){10}} \put(26,-6){$\sigma$} \put(70,1){=}
\put(87,3){\circle{6}} \put(84,1.5){\tiny$\times$}
\put(90,3){\line(1,0){40}} \put(133,3){\circle{6}}
\put(108,-6){$\sigma$} \put(85,8){$i$} \put(85,-8){$t$}
\put(132,-8){$t'$} \put(143,1){$+$}
\end{picture}
\nonumber \\ \nonumber \\ \nonumber \\
\begin{picture}(0,0)(77,0)
\put(160,3){\circle{6}} \put(158,8){$i$} \put(158,-8){$t$}
\put(179,-8){$t_1$} \put(206,-8){$t'$} \put(157,1.5){\tiny$\times$}
\put(163,3){\line(1,0){40}} \put(183,3){\circle{6}}
\put(183,12){\oval(13,16)[l]} \put(183,12){\oval(13,16)[r]}
\put(183,3){\color{white}\circle*{4.8}} \put(206,3){\circle{6}}
\put(168,-4){$\sigma$} \put(179,22){$\bar\sigma$}
\put(192,-4){$\sigma$} \put(216,1){$+$}
\end{picture}
\begin{picture}(0,0)(-150,0)
\put(4,3){\circle{6}} \put(2,8){$i$} \put(1,1.5){\tiny$\times$}
\put(7,3){\line(1,0){40}} \put(50,3){\circle{6}}
\put(27,5){\oval(16,13)[t]} \put(27,1){\oval(16,13)[b]}
\put(20,3){\circle{6}} \put(20,3){\color{white}\circle*{4.8}}
\put(35,3){\circle{6}} \put(35,3){\color{white}\circle*{4.8}}
\put(8,-4){$\sigma$} \put(24,14){$\bar\sigma$} \put(38,-4){$\sigma$}
\put(24,-15){$\bar\sigma$} \put(2,-8){$t$} \put(10,8){$t_1$}
\put(36,8){$t_2$} \put(50,-8){$t'$} \put(60,1){$+$}
\end{picture}
\nonumber \\ \nonumber \\ \nonumber \\ 
\begin{picture}(0,0)(-5,0)
\put(78,3){\circle{6}} \put(75,1.5){\tiny$\times$} \put(75,8){$i$}
\put(81,3){\line(1,0){40}} \put(100,12){\oval(13,16)[l]}
\put(100,12){\oval(13,16)[r]} \put(100,3){\circle{6}}
\put(100,3){\color{white}\circle*{4.8}} \put(128,5){\oval(16,13)[t]}
\put(128,1){\oval(16,13)[b]} \put(105,3){\line(1,0){38}}
\put(120,3){\circle{6}} \put(120,3){\color{white}\circle*{4.8}}
\put(136,3){\circle{6}} \put(136,3){\color{white}\circle*{4.8}}
\put(143,3){\line(1,0){10}} \put(154,3){\circle{6}}
\put(154,3){\color{white}\circle*{4.8}} \put(84,-4){$\sigma$}
\put(97,22){$\bar\sigma$} \put(108,-4){$\sigma$}
\put(126,4){$\sigma$} \put(126,-14){$\bar\sigma$}
\put(126,14){$\bar\sigma$} \put(140,-4){$\sigma$} \put(75,-8){$t$}
\put(95,-8){$t_1$} \put(111,8){$t_2$} \put(137,8){$t_3$}
\put(154,-8){$t'$}\put(170,-0){$+\cdots$}
\end{picture}
\label{eq16}
\end{eqnarray}
where $\buildrel i\over \otimes$ can be any site within the cluster
although we are particularly interested in the impurity site
$\put(4,2){\circle{6}}\quad$ or the site $\put(4,2){\circle*{6}}\quad$ at the edge of the cluster.
 We use the dressed one-particle
cluster Green function to incorporate all the diagrams of Eq.~({\ref{eq16}) into the expansion for the Green function
$G_{00,\sigma}(t-t')$, Eq.~(\ref{eq5}). This results in
\begin{widetext}
\begin{eqnarray}
&&\begin{picture}(0,0)(230,0) \put(0,0){\circle{6}}
\put(4,0){\line(1,0){10}} \put(13,-2){\rule{0.8cm}{2mm}}
\put(49,0){\circle{6}} \put(3,0){\line(1,0){10}}
\put(36,0){\line(1,0){10}} \put(-1,-14){$t$} \put(47,-14){$t'$}
\put(56,-2){$=$} \put(72,0){\circle{6}} \put(75,0){\line(1,0){10}}
\put(85,-2){\framebox(20,5)} \put(118,0){\circle{6}}
\put(105,0){\line(1,0){10}} \put(70,-14){$t$} \put(117,-14){$t'$}
\put(92,-14){$\sigma$} \put(126,-2){$+$} \put(141,0){\circle{6}}
\put(144,0){\line(1,0){10}} \put(154,-2){\framebox(20,5)}
\put(187,0){\circle*{6}} \put(174,0){\line(1,0){10}}
\put(187,0){\begin{xy}<1.4cm,0cm> \ar@{~}(2,0)
\end{xy}}
\put(216,0){\circle*{6}} \put(219,0){\line(1,0){10}}
\put(229,-2){\framebox(20,5)} \put(249,0){\line(1,0){10}}
\put(262,0){\circle{6}} \put(141,-14){$t$} \put(160,-14){$\sigma$}
\put(181,-14){$t_1$} \put(199,4){$\sigma$} \put(210,-14){$t_2$}
\put(238,-14){$\sigma$} \put(258,-14){$t'$} \put(270,-2){$+$}
\put(284,0){\circle{6}} \put(287,0){\line(1,0){10}}
\put(297,-2){\framebox(20,5)} \put(330,0){\circle{6}}
\put(317,0){\line(1,0){10}} \put(333,0){\line(1,0){10}}
\put(343,-2){\framebox(20,5)} \put(376,0){\circle{6}}
\put(363,0){\line(1,0){10}} \put(313,26){\begin{xy}<1.4cm,0cm>
\ar@{~}(2,0)
\end{xy}}
\put(327,29){$\bar\sigma$} \put(310,8){$\bar\sigma$}
\put(345,8){$\bar\sigma$} \put(306,32){$t_2$} \put(346,32){$t_3$}
\put(314,26){\circle*{6}} \put(346,26){\circle*{6}}
\put(329,2){\line(-2,3){14}} \put(346,23){\line(-2,-3){14}}
\put(316,18){\rotatebox{-53}{\rule{.58cm}{1.65mm}}}
\put(316.5,18){\color{white}\rotatebox{-53}{\rule{.55cm}{1.2mm}}}
\put(333,5){\rotatebox{56}{\rule{.58cm}{1.65mm}}}
\put(334,5.8){\color{white}\rotatebox{56}{\rule{.55cm}{1.2mm}}}
\put(282,-14){$t$} \put(304,-14){$\sigma$} \put(330,-14){$t_1$}
\put(350,-14){$\sigma$} \put(370,-14){$t'$} \put(385,-2){$+$}
\end{picture}
\nonumber\\ &&\nonumber\\
&&\nonumber\\
\begin{picture}(0,40)(230,0)
\put(0,0){\circle{6}} \put(0,-12){$t$} \put(3,0){\line(1,0){10}}
\put(13,-2){\framebox(20,5)} \put(18,-8){$\sigma$}
\put(33,0){\line(1,0){10}} \put(46.5,0){\line(0,1){15}}
\put(46.5,28){\line(0,1){15}} \put(47,42){\circle*{6}}
\put(35,43){$t_3$} \put(35,18){$\sigma$} \put(83,43){$t_4$}
\put(84,18){$\sigma$} \put(62,15){$\bar\sigma$}
\put(62,-5){$\bar\sigma$} \put(44,15){\framebox(5,13)}
\put(79,0){\line(0,1){15}} \put(79,28){\line(0,1){15}}
\put(79,42){\circle*{6}} \put(76.5,15){\framebox(5,13)}
\put(46,0){\circle{6}} \put(40,-12){$t'$}
\put(63,1){\oval(32,18)[t]} \put(63,-1){\oval(32,18)[b]}
\put(46,.1){\color{white}\circle*{4.8}}
\put(56,-12){\color{white}\rule{.55cm}{1.9mm}}
\put(56,-12){\framebox(15,5)}
\put(56,8){\color{white}\rule{.55cm}{1.9mm}}
\put(56,8){\framebox(15,5)} \put(79,0){\circle{6}}
\put(79,.1){\color{white}\circle*{4.8}}
\put(82,0){\line(1,0){10}}
\put(92,-2){\framebox(20,5)}
\put(112,0){\line(1,0){10}}
\put(126,-1){\circle{6}}
\put(126,-13){$t'$}
\put(102,-13){$\sigma$}
\put(45,40){\begin{xy}<1.4cm,0cm> \ar@{~}(2,0)
\end{xy}}
\end{picture}
\begin{picture}(0,40)(180,0)
\put(85,-2){$+$} \put(100,0){\circle{6}} \put(99,-12){$t$}
\put(122,-10){$\sigma$} \put(142,-12){$t_1$} \put(158,-10){$\sigma$}
\put(175,-12){$t_2$} \put(195,-12){$\sigma$} \put(215,-12){$t_3$}
\put(230,-10){$\sigma$} \put(252,-12){$t_4$} \put(270,-10){$\sigma$}
\put(297,-12){$t'$} \put(103,0){\line(1,0){10}}
\put(113,-2){\framebox(20,5)} \put(133,0){\line(1,0){10}}
\put(146,0){\circle*{6}} \put(146,0){\begin{xy}<1.4cm,0cm>
\ar@{~}(2,0)
\end{xy}}
\put(176,0){\circle*{6}} \put(179,0){\line(1,0){10}}
\put(189,-2){\framebox(20,5)} \put(209,0){\line(1,0){10}}
\put(222,0){\circle*{6}} \put(222,0){\begin{xy}<1.4cm,0cm>
\ar@{~}(2,0)
\end{xy}}
\put(252,0){\circle*{6}} \put(254,0){\line(1,0){10}}
\put(264,-2){\framebox(20,5)} \put(284,0){\line(1,0){10}}
\put(297,0){\circle{6}} \put(305,-2){$+\cdots$}
\end{picture}
\nonumber\\
\label{eq17}
\end{eqnarray}
\end{widetext}
After taking a Fourier transformation in time, we define the self-energies,
\newpage
\begin{eqnarray}
\nonumber\\
\Sigma^\sigma_{N}(\omega)&=&\hskip1.3cm
\begin{picture}(0,0)(45,0)
\put(9.5,2){\circle*{6}} \put(6,8){$V$} \put(21,5){$\sigma$}
\put(39,8){$V$} \put(10,2){\begin{xy}<1.4cm,0cm> \ar@{~}(2,0)
\end{xy}}
\put(40,2){\circle*{6}}
\end{picture}
=g_{N+1,\sigma}(\omega)V^2
\label{eq18a1}\\
\nonumber\\
\nonumber\\
&&\begin{picture}(0,0)(15,10) \put(14,25){\begin{xy}<1.4cm,0cm>
\ar@{~}(2,0)
\end{xy}}
\put(30,28){$\bar\sigma$} \put(11,5){$\bar\sigma$}
\put(44,5){$\bar\sigma$} \put(14,24){\circle*{6}}
\put(47,24){\circle*{6}} \put(29,0){\line(-2,3){14}}
\put(46,21){\line(-2,-3){14}}
\put(15,16){\rotatebox{-54}{\rule{.53cm}{1.68mm}}}
\put(15.5,16){\color{white}\rotatebox{-53}{\rule{.50cm}{1.2mm}}}
\put(33,3){\rotatebox{54}{\rule{.53cm}{1.68mm}}}
\put(34,3.7){\color{white}\rotatebox{54}{\rule{.50cm}{1.2mm}}}
\put(30,-3){\circle{6}} \put(58,-4){+} \put(85,26){\circle*{6}}
\put(85,-2){\line(0,1){26}} \put(85,26){\begin{xy}<1.4cm,0cm>
\ar@{~}(2,0)
\end{xy}}
\put(115,26){\circle*{6}} \put(115,-2){\line(0,1){26}}
\put(85,-2){\circle{6}} \put(100,-2){\oval(30,18)[t]}
\put(100,-2){\oval(30,18)[b]}
\put(85,-1.9){\color{white}\circle*{4.8}} \put(115,-2){\circle{6}}
\put(115,-1.9){\color{white}\circle*{4.8}}
\put(93,-13){\color{white}\rule{.55cm}{1.9mm}}
\put(93,-13){\framebox(15,5)}
\put(93,4){\color{white}\rule{.55cm}{1.9mm}}
\put(93,4){\framebox(15,5)}
\put(113,8){\color{white}\rule{1.7mm}{.45cm}}
\put(113,8){\framebox(5,13)}
\put(82,8){\color{white}\rule{1.7mm}{.45cm}}
\put(82,8){\framebox(5,13)} \put(98,28){$\bar\sigma$}
\put(74,11){$\bar\sigma$} \put(119,11){$\bar\sigma$}
\put(97,11){$\sigma$} \put(97,-20){$\sigma$} \put(133,-4){+}
\put(155,26){\circle*{6}} \put(155,-2){\line(0,1){26}}
\put(155,26){\begin{xy}<1.4cm,0cm> \ar@{~}(2,0)
\end{xy}}
\put(185,26){\circle*{6}} \put(185,-2){\line(0,1){26}}
\put(155,-2){\circle{6}} \put(170,-2){\oval(30,18)[t]}
\put(170,-2){\oval(30,18)[b]}
\put(155,-1.9){\color{white}\circle*{4.8}} \put(185,-2){\circle{6}}
\put(185,-1.9){\color{white}\circle*{4.8}}
\put(163,-13){\color{white}\rule{.55cm}{1.9mm}}
\put(163,-13){\framebox(15,5)}
\put(163,4){\color{white}\rule{.55cm}{1.9mm}}
\put(163,4){\framebox(15,5)}
\put(153,8){\color{white}\rule{1.7mm}{.45cm}}
\put(153,8){\framebox(5,13)}
\put(183,8){\color{white}\rule{1.7mm}{.45cm}}
\put(183,8){\framebox(5,13)} \put(166,28){$\sigma$}
\put(145,10){$\sigma$} \put(190,10){$\sigma$}
\put(167,11){$\bar\sigma$} \put(167,-22){$\bar\sigma$}
\put(197,-4){+}
\end{picture}\nonumber\\
\Sigma^\sigma_{0}(\omega)&=&\hskip6.9cm
\nonumber\\ \nonumber\\ \nonumber\\
&&\begin{picture}(0,0)(225,0) \put(229,20){\begin{xy}<1.4cm,0cm>
\ar@{~}(2,0)
\end{xy}}
\put(240,24){$\sigma$} \put(226,3){$\sigma$} \put(257,3){$\sigma$}
\put(229,20){\circle*{6}} \put(262,20){\circle*{6}}
\put(270,-8){$+\cdots$} \put(245,-5){\line(-2,3){16}}
\put(260,17.5){\line(-2,-3){16}} \put(245,-6){\circle{6}}
\put(245,-17){\oval(20,24)[t]} \put(245,-17){\oval(20,24)[b]}
\put(234,-23){\color{white}\rule{1.5mm}{.42cm}}
\put(234,-23){\framebox(4,12)}
\put(252,-23){\color{white}\rule{1.5mm}{.42cm}}
\put(252,-23){\framebox(4,12)}
\put(245,-6){\color{white}\circle*{4.86}}
\put(245,-29){\color{white}\circle*{4.86}} \put(245,-28){\circle{6}}
\put(227,-18){$\bar\sigma$} \put(257,-18){$\bar\sigma$}
\put(230,13){\rotatebox{-54}{\rule{.53cm}{1.68mm}}}
\put(230.8,13){\color{white}\rotatebox{-54}{\rule{.50cm}{1.2mm}}}
\put(248,1){\rotatebox{54}{\rule{.53cm}{1.68mm}}}
\put(248.7,1.6){\color{white}\rotatebox{54}{\rule{.50cm}{1.2mm}}}
\end{picture}
\nonumber\\ \nonumber\\
\label{eq18a}
\end{eqnarray}
where we have explicitly drawn the contribution to the self-energy
up to terms proportional to $U^2$.

The Green function of the system at the impurity
can be written as a general Dyson equation:
\begin{equation}
G_{00,\sigma}(\omega)=G^{(0)}_{00,\sigma}(\omega)+\sum_{i}
G^{(0)}_{0i,\sigma}(\omega)\Sigma^\sigma_{i}(\omega)G_{i0,\sigma}
(\omega) \label{18b}
\end{equation}
where $i$ is restricted to be either $0$ or $N$ and the self-energy
$\Sigma^\sigma_{i}(\omega)$ is defined as
\begin{equation}
\Sigma^\sigma_{i}(\omega)=\Sigma^\sigma_{N}(\omega)\delta_{iN}+\Sigma^\sigma_{0}(\omega)\delta_{i0}.
\label{eq18c}
\end{equation}
In order to compare the relative contribution of 
$\Sigma_0^\sigma$ and $\Sigma_N^\sigma$, which crucially depends upon
the cluster size N, we proceed as follows. Considering that $V=t_N$
and using Eqs. (\ref{eq18a1}) and (\ref{eq19}) we have that
\begin{equation}
\Sigma_N^\sigma(\omega) \propto V^2 \sim \lambda^{-(N-1)},
\end{equation}
where we have ignored the Green function $g_{N+1,\sigma}(w)$ because
we have numerically verified that in the neighborhood of the Fermi
energy this function is independent of N.

To evaluate the dependence of $\Sigma_0^\sigma$ upon N, we observe
that all terms in Eq. (\ref{eq18a}) are multiplied by the square of
the non-diagonal cluster propagator 
\begin{equation}
G^{(0)}_{0 N+1}=
\begin{picture}(0,0)(3,0) \put(8,4){\line(1,0){10}}
\put(5,4){\circle{6}} \put(18,2){\framebox(26,5)}
\put(44,4){\line(1,0){10}} \put(57,4){\circle*{6}} \put(3,-9){$^t$}
\put(29,-7){$^{\bar{\sigma}}$} \put(55,-9){$^{t'}$}
\end{picture}\hskip2cm \,. 
\end{equation}
In addition, the dependence of this propagator on $\lambda$ is given by 
\begin{equation}
M(\omega)\Pi_{i=0}^N t_i \sim \lambda^{-(N-1)N/4}, 
\end{equation}
where the function $M(\omega)$ goes asymptotically
to zero when N increases above a characteristic length, which in our
case corresponds to the size of the Kondo cloud. Defining
$f(N)=M^2(\omega)$ we obtain,

\begin{equation}
\Sigma_0^\sigma(\omega) \sim f(N) ~ \lambda^{-(N-1)N/2}.
\end{equation}

As discussed in the main text in Sec.~\ref{sec:ldeca}, the
contribution to the self-energy $\Sigma^\sigma_{0}(\omega)$ can be
neglected when compared with $\Sigma^\sigma_{N}(\omega)$ when the density of states 
of the leads is logarithmically discretized, as their ratio is then proportional to:

\begin{equation}
\frac{\Sigma^\sigma_{0}(\omega)}{\Sigma^\sigma_{N}(\omega)}  \sim f(N) ~\lambda^{-(N-1)(N/2-1)}\,. \label{eq20}
\end{equation}
In this case, the embedding process is extremely simplified.


\begin{thebibliography}{99}

\bibitem{molecular} C. Joachim, J. K. Gimzewski, and A. Aviram,
Nature {\bf 408}, 541 (2000).

\bibitem{qcomp} F. H. L. Koppens,
C. Buizert, K. J. Tielrooij, I. T. Vink, K. C. Nowack, T. Meunier,
L. P. Kouwenhoven,  and L. M. K. Vandersypen,
Nature {\bf 442}, 766 (2006).

\bibitem{GG} D.~Goldhaber-Gordon,
H. Shtrikman, D. Mahalu, D. Abusch-Magder, U. Meirav, and M. A. Kastner,
\newblock Nature {\bf 391}, 156 (1998).

\bibitem{molecules} J. Park,
A.N. Pasupathy, J.I. Goldsmith, C. Chang, Y. Yaish, J.R. Petta,
M. Rinkoski, J. P. Sethna, H. D. Abruna, P. L. McEuen and D. C. Ralph,
Nature {\bf 417}, 722 (2002).

\bibitem{nonfermi} R. M. Potok,
I. G. Rau, H. Shtrikman, Y. Oreg,  and D. Goldhaber-Gordon,
Nature {\bf 446}, 167 (2006).

\bibitem{bethe} N. Andrei, Phys. Rev. Lett. {\bf 45}, 379 (1980);
J. Bon\v{c}a,  A. Ram\v{s}ak and T. Rejec, cond-mat/0407590v2.

\bibitem{Wilson} K. G. Wilson, Rev. Mod. Phys. {\bf 47}, 773 (1975).
For a recent review, see: R. Bulla, T. Costi, and T. Pruschke, Rev. Mod. Phys. {\bf 80}, 395 (2008).

\bibitem{tDMRG} S. R. White and A.E. Feiguin, Phys. Rev. Lett. {\bf
 93}, 076401 (2004);
 A. Daley,
 C. Kollath, U. Schollw\"ock, and G. Vidal,
 J. Stat. Mech.: Theory Exp., P04005 (2004);
 K. Al-Hassanieh,
 A. E. Feiguin, J. A. Riera, C. A. B\"usser, and E. Dagotto,
 Phys. Rev. B {\bf 73}, 195304 (2006).

\bibitem{fRG} C. Karrasch,
T. Enss, and V. Meden,
Phys. Rev. B {\bf 73}, 235337 (2006).

\bibitem{ferrari99} V. Ferrari,
G. Chiappe, E. V. Anda, and M. A. Davidovich,
Phys. Rev. Lett. {\bf 82}, 5088 (1999).

\bibitem{davidovich02} M. A. Davidovich,
E. V. Anda, C. A. B\"usser, and G. Chiappe,
Phys. Rev. B {\bf 65}, 233310 (2002).

\bibitem{anda02} E. V. Anda,
C. A. B\"usser, G. Chiappe, and M. A. Davidovich,
Phys. Rev. B {\bf 66}, 035307 (2002).

\bibitem{chiappe03} G. Chiappe and J. A. Verges, J. Phys.: Cond. Matt. {\bf 15}, 8805 (2003).

\bibitem{busser2} C. A. B\"usser, E. V. Anda, L. Urba, V. Ferrari, G. Chiappe, J. Magn.
Magn. Mater. {\bf 177-181}, 311 (1998).

\bibitem{chiappe99} G. Chiappe, C. A. B\"usser, E. V. Anda, V. Ferrari,
J. Phys.: Cond. Matter {\bf 27}, 5237 (1999).

\bibitem{Tremblay}
S. Psirsult, D. S\'en\'echal, and A.-M. S. Tremblay, Eur. Phys. J. B {\bf 16}, 85 (2000);
D. S\'en\'echal, D. Perez, and M. Pioro-Ladri\`ere, Phys. Rev. Lett. {\bf 84}, 522 (2000);
D. S\'en\'echal, D. Perez, and D. Plouffe, Phys. Rev. B {\bf 66}, 075129 (2002).

\bibitem{white} S. R. White, Phys. Rev. Lett. {\bf 69}, 2863 (1992);
U. Schollw\"ock, Rev. Mod. Phys. {\bf 77}, 259 (2005); K. Hallberg, Adv. Phys. {\bf 55} 477
 (2006).

\bibitem{hofstetter} W. Hofstetter, Phys. Rev. Lett. {\bf 85}, 1508 (2002);
F. Verstraete, A. Weichselbaum, U. Schollw\"ock, J. I. Cirac, and J. von Delft,
cond-mat/0504305.

\bibitem{anders} F. Anders and A. Schiller, Phys. Rev. Lett. {\bf 95},
 196801 (2005).

\bibitem{busser04} C. A. B\"usser, A. Moreo, and E. Dagotto, 
Phys. Rev. B {\bf 70}, 035402 (2004).

\bibitem{hm08} See: F. Heidrich-Meisner,
G. B. Martins, C. A. B\"usser, K. A. Al-Hassanieh, A. E. Feiguin, G. Chiappe, E. V. Anda and E. Dagotto,
arxiv:0705.1801 for an extended discussion.

\bibitem{su4} P. Jarillo-Herrero, J. Kong Herre, S. J. van der Zant, C. Dekker, L. P. Kouwenhoven, and S. De Franceschi,
Nature {\bf 434}, 484 (2005); A. Makarovski, J. Liu, and G. Finkelstein, Phys. Rev. Lett. {\bf 99}, 066801 (2007).

\bibitem{ecasu4}
C. A. B\"usser and G. B. Martins, Phys. Rev. B {\bf 75}, 045406 (2007).

\bibitem{wiel} W. G. van der Wiel, S. De Franceschi, J. M. Elzerman, 
S. Tarucha, L. P. Kouwenhoven, J. Motohisa, F. Nakajima, and T. Fukui, 
Phys. Rev. Lett. {\bf 88}, 126803 (2002).

\bibitem{two-stage}  M. Vojta, R. Bulla, and W. Hofstetter, Phys. Rev. B {\bf 65}, 140405 (2005). 

\bibitem{Grempel} P. S. Cornaglia and G. Grempel, Phys. Rev. B {\bf 71}, 075305 (2005).

\bibitem{pedro}	G. A. Lara, P. A. Orellana, J. M. Yanez, and E. V. Anda, Sol. Stat. Comm. {\bf 136}, 323 (2005).

\bibitem{zitko06}R. {\v{Z}}itko and J. Bon{\v{c}}a, Phys. Rev. B {\bf 73}, 035332 (2006).

\bibitem{zitko07}R. {\v{Z}}itko and J. Bon{\v{c}}a, Phys. Rev. Lett. {\bf 98}, 047203 (2007).

\bibitem{quique} A. Zhao, Qunxiang Li, L. Chen, H. Xiang, W. Wang, Sh. Pan, B. Wang,
X. Xiao, J. Yang, J. G. Hou, and Q. Zhu, Science {\bf 309}, 1542 (2005);
G. Chiappe, and E. Louis, Phys. Rev. Lett. {\bf 97}, 076806 (2006);
J. M. Aguiar-Hualde, G. Chiappe, E. Louis, and E. V. Anda, Phys. Rev. B {\bf 76}, 155427 (2007).

\bibitem{martins05} G. B. Martins, C. A. B\"usser, K. A. Al-Hassanieh, A. Moreo, and E. Dagotto,
Phys. Rev. Lett. {\bf 94}, 026804 (2005).

\bibitem{interference} C. A. B\"usser, G. B. Martins, K. A. Al-Hassanieh,
A. Moreo, and E. Dagotto, Phys. Rev. B {\bf 70}, 245303 (2004).

\bibitem{otras1} G. Chiappe and A. A. Aligia, Phys. Rev. B {\bf 66}, 075421 (2002).

\bibitem{otras2} M. E. Torio, K. Hallberg,
 A.H. Ceccatto, and C. R. Proetto, Phys. Rev. B {\bf 65}, 085302 (2002).

\bibitem{Armando} A. A. Aligia and A. M. Lobos, J. Phys.: Condens. Matter {\bf 17}, S1095 (2005).

\bibitem{Hewsonbook} A. C. Hewson, {\it The Kondo Problem to Heavy
Fermions} (Cambridge University Press, 1997).

\bibitem{Elbio} E. Dagotto, Rev. Mod. Phys. {\bf 66}, 763 (1994).

\bibitem{Abrikosov} A. A.~Abrikosov, L. P.~Gorkov, and I. E. Dzyaloshinski,
{\it Methods of Quantum Field Theory in Statistical Physics} (Dover Books, 1963).

\bibitem{qual} E. M\"uller-Hartmann, Z. Phys. B-Cond. Matt. {\bf 76}, 21 (1989).

\bibitem{hewson2} See discussion on Section 4.3 of reference \onlinecite{Hewsonbook} (pgs. 78 -- 81).

\bibitem{Affleck} P. Simon and I. Affleck, Phys. Rev. B {\bf 68}, 115304 (2003); 
L. Borda, Phys. Rev. B {\bf 75}, 041307 (2007); 
J. Simonin, arXiv:0708.3604.

\bibitem{Busser1} C.A. B\"usser, 
E. V. Anda, A. L. Lima, M. A. Davidovich, and G. Chiappe,
Phys. Rev. B {\bf 62}, 9907 (2000).

\bibitem{conduc} Y. Meir, N. S. Wingreen, and P. A. Lee, Phys. Rev. Lett. {\bf
66}, 3048 (1991); E. V. Anda and F. Flores, J. Phys.: Condens. Matter
{\bf 3}, 9087 (1991).

\bibitem{bulla} R. Bulla, T. A. Costi, and D. Vollhardt, Phys. Rev. B {\bf 64}, 045103 (2001);
O. Sakai, Y. Shimizu and T. Kasuya, J. Phys. Soc. Jpn. {\bf 58}, 3666 (1989);
T. A. Costi, A. C. Hewson, and Zlati\'c, J. Phys.: Condens. Matter {\bf 6}, 2519 (1994).

\bibitem{Coqblin} B. Coqblin, J. Arispe, J. R. Iglesias, C. Lacroix and K. LeHur,
J. Phys. Soc. Jpn. {\bf 65}, Suppl. B, 64 (1996).

\bibitem{note-inter} The reader is reminded that an alternate interpretation 
of the conductance suppression involves the destructive interference between the two 
paths available to the conduction electrons: a path that avoids the dots 
and a path that visits them. This interpretation is completely equivalent 
to the back-scattering one. These two interpretations will be used interchangeably 
along the text.

\bibitem{Guille} G. Chiappe, J. Fern\'andez-Rossier, E. Louis, and E. V. Anda,
Phys. Rev. B. {\bf 72}, 245311 (2005)

\bibitem{metzner}  
E. V. Anda, J. Phys. C-Solid State Phys. {\bf 14 }, L1037 (1981); 
W. Metzner, Phys. Rev. B {\bf 43}, 8549 (1991).

\end{thebibliography}
\end{document}